# Plasmonic modes of polygonal particles calculated using a quantum hydrodynamics method

Kun Ding and C. T. Chan[†]

*Department of Physics and Institute for Advanced Study, The Hong Kong University of Science and Technology, Clear Water Bay, Hong Kong*

[†] Corresponding E-mail: phchan@ust.hk

**Abstract**

Plasmonic resonances of nanoparticles have drawn lots of attentions due to their interesting and useful properties such as strong field enhancements. These systems are typically studied using either classical electrodynamics or fully quantum theory. Each approach can handle some aspects of plasmonic systems accurately and efficiently, while having its own limitation. The self-consistent hydrodynamics model has the advantage that it can incorporate the quantum effect of the electron gas into classical electrodynamics in a consistent way. We use the method to study the plasmonic response of polygonal particles under the influence of an external electromagnetic wave, and we pay particular attention to the size and shape of the particle and the effect of charging. We find that the particles support edge modes, face modes and hybrid modes. The charges induced by the external field in the edge (face) modes mainly localize at the edges (faces), while the induced charges in the hybrid modes are distributed nearly evenly in both the edges and faces. The edge modes are less sensitive to particle size than the face modes, but are sensitive to the corner angles of the edges. When the number of sides of regular polygons increases, the edge and face modes gradually change into the classical dipole plasmonic mode of a cylinder. The hybrid modes are found to be the precursor of the Bennett mode, which cannot be found in classical electrodynamics.

**PACS numbers (PhySH):**

**Section I. Introduction**

Plasmonic resonances are ubiquitous for metallic nanoparticles [1-8] and these collective excitation modes can be used to realize many interesting phenomena, such as extraordinary optical transmission [7] and negative refraction [8]. Recent advances in nanoscience enabled the fabrication of metallic nanoparticles with controlled size and shape. At the nanoscale, the quantum nature of electrons emerges [9-26] and standard local electrodynamics description becomes inadequate. For example, nonlocal effects must be taken into considerations in order to study the resonances of nano metallic particles, as have already been verified experimentally [9,10]. In addition to the nonlocal effect, other quantum effects are also not ignorable in plasmonic modes [11-26]. One issue is the tunneling of the electrons through tiny gaps for gap plasmonic modes in the core shell structure [11-13] and dimer structures [14-17], which reduces the field enhancement. Another quantum feature is the loss caused by the surrounding material due to the creation of surface electron-hole pairs [25,26]. In order to describe these quantum effects theoretically, different methods have been proposed [27-50]. The most accurate formulation that is computationally tractable is perhaps the time-dependent density functional theory (TD-DFT) [27-33]. This method is computationally intensive, and it is not that easy to couple TDDFT with Maxwell equations to include electromagnetic (EM) wave characteristics, such as retardation. On the other hand, corrected classical models, such as nonlocal surface layer models [34-36], hard-wall hydrodynamics model [37-40], quantum corrected model [41-43] and others [44,45], have also been proposed to study these systems. These models can handle full electromagnetic wave characteristics, in the sense that Maxwell equations are explicitly and exactly solved, but some quantum features of the electron gas are not included. For example, the electron spill out effect near the metal surface [46] is typically not treated adequately. The self-consistent hydrodynamics model [46-50] can be viewed as an advanced form of quantum corrected classical model. This model works with the electron density instead of electronic wave functions, so that the energy functional is simple enough to solve for the response equations and Maxwell equations at the same time and on the same footing self-consistently [46-50]. This method has been employed successfully to study the plasmonic responses for some simple plasmonic structures with simple geometries [46-50]. But the power of this method is its ability to handle systems with complex structure and complex geometry and it is worthwhile to study the EM wave response of some complicated metallic particles using this model.

In this paper, we employ the self-consistent hydrodynamics model to investigate the plasmonic modes of polygonal particles. We found three types of plasmonic modes, namely edge, face and hybrid modes. In the edge (face) modes, the charges induced by an external EM wave are mainly localized at the edges (faces), while in the hybrid modes, the induced electrons are distributed nearly evenly in both edges and faces. The resonance frequency of the edge modes are mainly determined by the corner angles of the edges, while the face modes are more sensitive to particle sizes. In addition, when the number of sides of regular polygons increases, the edge and face modes merge into the classical dipole plasmonic mode of a cylinder. The hybrid modes are found to be the precursor of the Bennett mode, which originates from charge oscillating in and out of a surface. The paper is organized as follows. In Section II, we briefly review the self-consistent hydrodynamics model. The numerical results of ground states for charge neutral particles are given in Section III. In Section IV, we calculate the scattering properties of these polygonal particles, such as absorption spectrum, plasmonic mode profile, geometric effect, energy dissipation distributions, and the effect of charging. Conclusions are drawn in Section VI.

**Section II. Formulation of self-consistent hydrodynamics model**

The systems studied are two dimensional (2D) polygonal particles, as shown in Fig. 1(a). The yellow regions stand for the jellium background in the shape of regular polygons. They are bounded by a circumscribed circle with radius $r_a$, as marked by the gray dashed lines in Fig. 1(a). The right inset in Fig. 1(a) shows polygonal particles with different side numbers $N$. The blue dashed circle and triangle show the boundaries of ground state and scattering state calculation domain, respectively, denoted by $\Omega_G$ and $\Omega_D$.

To study the plasmonic response of our system, we employ a self-consistent hydrodynamics model (SC-HDM) [46-50]. The SC-HDM is well documented in the literature [46-50], but to make the descriptions self-contained, we will give a short introduction of the model in this section. The basic assumption of hydrodynamics model (HDM) is to treat the electrons of metals as ideal, irrotational, and isentropic fluid which indicates that temperature effects are not included in the model. Based on these assumptions, the electron gas can be described by the physical variables, electron density $n(\mathbf{r},t)$, and canonical momentum $\mathbf{p}(\mathbf{r},t)$. The equations of motion for the electron gas can be obtained from the time evolutions of these variables, namely [46-50],

$$m_e n \left( \frac{\partial \mathbf{v}}{\partial t} + \mathbf{v}\cdot\nabla\mathbf{v} \right) = -n\nabla\left( \frac{\delta G}{\delta n} \right) + nq_e(\mathbf{E}+\mathbf{v}\times\mathbf{B}), \qquad (2.1)$$

$$\frac{\partial n}{\partial t} + \nabla\cdot(n\mathbf{v}) = 0, \qquad (2.2)$$

where $q_e = -e$ is the charge of electron $(e>0)$, $m_e$ is the mass of electron, $\mathbf{v} = (\mathbf{p}-q_e\mathbf{A})/m_e$ is the velocity of electron, $\mathbf{A}$ is the vector potential, and $\mathbf{E},\mathbf{B}$ are the fields generated by the charges, currents and external electromagnetic waves. The functional $G[n(\mathbf{r},t)]$ is the internal energy of the electron gas, including the internal kinetic energy and the exchange-correlation energy [51-53]. In this work, we follow the choice of $G[n(\mathbf{r},t)]$ in Ref. [47] (see also Supplemental Material, Sec. I [54]). Equations (2.1)-(2.2) are the basic equations in the hydrodynamics model, and their numerical solutions together with the Maxwell equations are needed to study the plasmonic behavior of our system.

The next step is to do linear expansions of the following quantities in Eqs. (2.1)-(2.2) as

$$n = n_0 + n_1, \qquad \mathbf{v} = \mathbf{v}_1,$$
$$\left(\frac{\delta G}{\delta n}\right) = \left(\frac{\delta G}{\delta n}\right)_0 + \left(\frac{\delta G}{\delta n}\right)_1, \qquad (2.3)$$
$$\mathbf{E} = \mathbf{E}_0 + \mathbf{E}_1, \qquad \mathbf{B} = \mathbf{B}_1,$$

where the subscript "0" stands for the equilibrium states and subscript "1" means the linear expansions of these quantities. The equilibrium $\mathbf{v}_0$ is zero which means no electrons move at equilibrium, and $\mathbf{B}_0$ is also zero because no external static magnetic field is imposed. Substitute the expansions (2.3) into Eqs. (2.1)-(2.2) and Maxwell equations, and keep the zero order terms, then we arrive at

$$\left(\frac{\delta G}{\delta n}\right)_0 + q_e \phi_0 = \mu, \qquad (2.4)$$

$$\nabla^2 \phi_0 = \frac{q_e}{\varepsilon_0}(n_+ - n_0), \qquad (2.5)$$

where $\mu$ is the chemical potential, $n_0$ is ground state electron density, and $\phi_0$ is electrostatic potential defined as $\mathbf{E}_0 = -\nabla \phi_0$. $n_+$ stands for the positive charged background including nuclei and core electrons. Here, we use the jellium model $n_+ = n_j$, in which $n_j = n_{ion}$ inside the metal, and $n_j = 0$ outside the metal. The ion density $n_{ion}$ is defined via dimensionless quantity $r_s$ as

$$n_{ion} = \frac{3}{4\pi(r_s a_H)^3}, \qquad (2.6)$$

where $a_H = 0.529\text{Å}$ is Bohr radius. The total number of electrons imposes another constraint [46,55]. Suppose the particle has a net charge $\sigma$, then the electron density should obey

$$\int_{\Omega_G} d\mathbf{r}\, e[n_+ - n_0] = \sigma. \qquad (2.7)$$

Equations (2.4), (2.5) and (2.7) collectively determine the ground state charge densities of the plasmonic particles.

The linear response equations in the frequency domain could be obtained by keeping the first order terms in the expansions of Eqs. (2.1)-(2.2) and Maxwell equations, namely [46-50]

$$(-i\omega+\gamma)\mathbf{J}_1 = \frac{en_0}{m_e}\nabla\left(\frac{\delta G}{\delta n}\right)_1 + \frac{e^2 n_0}{m_e}\mathbf{E}_1, \qquad (2.8)$$

$$\nabla\cdot\mathbf{J}_1 - i\omega\rho_1 = 0, \qquad (2.9)$$

$$\nabla\times(\nabla\times\mathbf{E}_1) - \left(\frac{\omega}{c}\right)^2\mathbf{E}_1 = i\omega\mu_0\mathbf{J}_1, \qquad (2.10)$$

where the induced currents and charges are defined as $\mathbf{J}_1 = n_0 q_e \mathbf{v}_1$ and $\rho_1 = q_e n_1$, $\gamma$ is the loss parameter introduced empirically here [56,57], and the electric fields $\mathbf{E}_1 = \mathbf{E}_{inc} + \mathbf{E}_{sca}$ are total fields including incident fields and scattering fields from the environment. Equations (2.8) to (2.10) determine the linear responses of the plasmonic particles.

Two numerical steps are needed to obtain the plasmonic resonances and related properties of the system using SC-HDM. We first solve Eqs. (2.4), (2.5) and (2.7) to get the ground states, and then solve Eqs. (2.8) to (2.10) to get the scattering states. In the following sections, we will discuss the numerical implementations of the models and some numerical results of polygonal particles.

**Section III. Ground state of charge density of neutral polygonal particles**

In this section, we consider the ground state density for the charge neutral polygonal particles. We employ a finite element method (FEM) to solve the differential equations (2.4), (2.5), and (2.7) in real space [58,59], and numerical details are given in Supplemental Material, Sec. II [54]. For simplicity's sake, we will assume throughout this work that the jellium represents the simple metal sodium. In the numerical calculations, the ground state calculation domain is denoted by $\Omega_G$, the boundary of which is shown by blue dash lines in Fig. 1(a). In the energy functional, the parameter $\lambda_\omega$ comes from von Weizsacker kinetic energy functional [51,52]. Throughout this work, we set $\lambda_\omega$ to 0.12 as this is optimal for our calculations (see Supplemental Material, Sec. IV [54]).

The calculated electron density $n_0$ distributions of ground state for a sodium ($r_s = 4$) triangle with radius $r_a = 2\text{nm}$ is shown in Fig. 1(b). To show the difference between electron density and jellium background, we plot the relative dimensionless quantity $(n_0 - n_j)/n_{\text{ion}}$ in Fig. 1(b) to demonstrate the ground state charge densities. The white dashed line shows the boundaries of jellium background. We note that the electron density distributions are smooth near the edges of the jellium background, although the jellium background density does have singularities at the polygon corners. As expected, the electron density is equal to the jellium background in the bulk metal region. The difference between electron density and jellium background only occurs near the surfaces, creating a surface dipole layer. The existence of this dipole layer forms an electrostatic potential barrier, which constitutes one part of the effective single electron potential $V_{\text{eff}}(\mathbf{r})$ (see Supplemental Material, Sec. II, for the explicit formulations of $V_{\text{eff}}(\mathbf{r})$ [54]). To see this, we plot the corresponding effective single electron potential $V_{\text{eff}}(\mathbf{r})$ in Fig. 1(c). The zeros of the effective potential are set to the chemical potential deep inside the metal. The distribution of $V_{\text{eff}}(\mathbf{r})$ is anisotropic, indicating that the surface dipole layer in the edge direction (E-direction), denoted by green arrow in Fig. 1(a), are quite different from that in the face direction (F-direction), labeled by the red arrow in Fig. 1(a). It should be mentioned that our calculations are in two dimensional (2D) space, with the z-direction being invariant, so the corner in 2D space is the edge in the three dimensional (3D) space, and the edge in 2D space is the face in 3D space. Throughout this work, we will use the notations in 3D space.

To further see the effect of the sharp edges on the surface dipole, we plot the modulation of the local work function as a function of polar angles $\theta/\pi$ for a sodium triangle by gray lines in Fig. 2(a). The local work function $W_F(\theta)$ is defined as $W_F(\theta) = V_{eff}(\theta, r = r_g) - V_{eff}(\theta, r = 0)$, which $r$ and $\theta$ are radial and angular coordinate in the polar coordinate system. Since the cylinder is isotropic, so we choose the work function of a sodium cylinder as the reference, marked by magenta lines in Fig. 2(a). The gray line in Fig. 2(a) shows that the local work function in the E-direction is lower than that in the F-direction. For comparison, we also plot the local work function distributions for a sodium square, pentagon, hexagon, and 20-sided polygon in Fig. 2(a) by orange, red, blue, and green lines, respectively. As the number of sides of the regular polygon increases, the local work function difference between polygons and cylinder becomes smaller because the corners of the polygon are smoother. The local work function difference between E-direction and F-direction is also smaller with the increasing of side number. To show this, we plot the relative local work functions versus the half of the interior angle ($\alpha = \frac{\pi}{2}\left(1 - \frac{2}{N}\right)$) of a regular polygon in Fig. 2(b) for the E-direction and F-direction by open squares and open triangles, respectively. It can be seen that as $\alpha$ increases, the difference between the edge and face direction gets smaller. As far as the ground state is concerned, the key difference between polygons and the circle is the anisotropic local work function, indicating different surface dipoles in different angles. This is a geometric curvature effect, in contrast to the crystal plane (packing) induced work function difference [60].

**Section IV. Scattering properties of polygonal particles**

After obtaining the ground state density, we can implement the linear response equations into FEM. The numerical details are given in Supplemental Material, Sec. III [54]. To get the scattering state accurately, the total calculation domain ($\Omega_T$) must be of the wavelength order. As we are concerned with the deep subwavelength properties of the plasmonic resonances, the particle size is much smaller than the wavelength. In order to deal with the scale difference between the particle size and the wavelength, we divide the total calculation domain into two subdomains. One is the domain near the particle denoted by $\Omega_D$ within which we solve the full linear response equations, and the other subdomain $\Omega_F \left( \equiv \Omega_T - \Omega_D \right)$ is the free space within which we only solve Maxwell equations. The subdomain $\Omega_D$, with a finer grid than $\Omega_F$, is chosen to make sure that there are no induced electrons outside this domain, so we do not need to solve linear response equations outside. Throughout this work, we set the radius $r_d$ of $\Omega_D$ to $r_d = r_a + 0.7\text{nm}$, and this setting is optimal for our calculations (see Supplemental Material, Sec. V [54]).

**A. Absorption spectrum and mode profile**

To see the plasmonic modes of polygonal particles, we calculate the absorption cross section $\sigma_{abs}$ under plane wave illumination with incident angle $\theta_i = 0$ ($\mathbf{E}//\mathbf{y}$) for the triangle and square, as shown in Figs. 3(a) and 3(b) by solid blue/red lines, respectively. The incident angle $\theta_i$ is defined as the polar angle of the wave vector, and the coordinates are given in Fig. 1(a). The values of $\sigma_{abs}$ are calculated by integrating the Poynting vector in the far field. From Figs. 3(a) and 3(b), several prominent absorption peaks are found in the spectra for both the triangle and square. The frequency of resonance peaks shows strong dependence on the geometry. If we characterize the resonances according to the pattern of induced charge concentration, we can group them into three categories: edge modes (Ed-mode), face modes (Fa-mode), and hybrid modes (Hb-mode). The edge (face) modes have the induced electrons mainly localize at the edges (faces). In the hybrid modes, the induced electrons are distributed nearly evenly in both edges and faces. Using this classification, we labeled the plasmonic modes by *Ed*, *Fa*, and *Hb* in Fig. 3. For instance, the symbol $\text{Ed}_1^\text{T}$ is used to denote the first (subscript "*1*" is the mode index) edge ($Ed$) mode for triangle (superscript "$T$") particles. The lowest frequency modes for both the triangle and square are found to be edge modes. The induced charge and current distributions of the $Ed_1^T$ mode and $Ed_1^S$ mode are plotted in Figs. 4(a) and 4(c), respectively. The color stand for the quantity $\rho_1/E_0$ and the arrows represent $\mathbf{J}_1$ at some particular time. It is clear that the induced electrons are mainly concentrated near the edges. There are also high order edge modes, such as $Ed_2^T$. The induced electrons in these modes still mainly localize near the edges, but they decay into the face slower than the fundamental mode.

The face modes for triangle and square appear in higher frequencies, as shown by $Fa_1^T$ and $Fa_1^S$ in Figs. 3(a) and 3(b). The induced charge and current distributions of $Fa_1^T$ and $Fa_1^S$ modes are shown in Figs. 5(a) and 5(c), respectively. Compared with the Ed-modes, Fa-modes have their induced electrons mainly localized near the polygon faces, with the currents flowing from one face to another. The induced charges have opposite signs for a single face, but the amplitudes of the positive and negative charges are different. In Figs. 3(a) and 3(b), the further higher frequency modes are hybrid modes, denoted by $Hb^T$ and $Hb^S$. The bandwidths of them are broad due to more dissipative loss, and the induced charges are in both edges and faces.

For comparison, solid gray lines in Figs. 3(a) and 3(b) show the absorption spectrum of the triangle and square calculated by the classical model. The classical model is equivalent to the SC-HDM under local response approximations (LRA, see also Supplemental Material, Sec. VI [54]), which ignores the $\frac{\delta G}{\delta n}$ term and sets $n_0 = n_j$. We also see several prominent absorption peaks in the spectrum. The modes could be classified into two classes. The lower frequency peaks (below 4eV) are edge modes, while the higher frequency peaks can be assigned as the face modes. This classification is possible because there are two branches of plasmonic modes for a single metallic wedge according to prior quasi-static solutions of the wedge problem [61,62]. The lower frequency band has even symmetry, while the higher frequency band consists of odd modes. For the even mode, the induced charges on the two faces of the wedge are the same, so the induced charges are maximal in the corner position. Symmetry-wise, this corresponds to the edge modes observed in SC-HDM. For the odd modes, the induced charges of the two faces have opposite signs, so there must be a node in the corner position. This corresponds to the face modes observed in SC-HDM. These classical results help us understand the modes in SC-HDM qualitatively, but they are significant shifting in frequency due to quantum spill out effects [63]. We will discuss the physical reasons for the quantum corrections later.

We now change the incident angle $\theta_i$ of the plane wave but keep the incident electric fields in the xy plane, and then calculate absorption spectrum again. For triangle particles, we plot the absorption spectrum for plane wave with incident angle $\theta_i = \pi/2$ in Fig. 3(a) by open gray dots. We see that the absorption spectrum is almost the same as the spectrum for $\theta_i = 0$ case. This is because the particle size is much smaller than the wavelength, and as such, the EM wave cannot recognize the anisotropy of the metallic particles [64] and hence the external wave couples equally well with the particle resonances irrespective of the incident angle. We plot the induced charges and currents under this plane wave incidence of $Ed_1^T$ and $Fa_1^T$ modes in Figs. 4(b) and 5(b), respectively. The induced electrons are still localized in edges and faces for these modes even though the polarizations of the incident waves are different. These results further show that the peaks correspond to intrinsic resonant modes. For the square particle, the absorption spectrum under plane wave illumination ($\theta_i = \pi/4$) is also plotted by green dots in Fig. 3(b). The induced charges and currents of $Ed_1^S$ and $Fa_1^S$ modes are plotted in Figs. 4(d) and 5(d), respectively. We note that these edge and face modes also exist for various geometric configurations, and they are experimentally observed in different plasmonic particles

[1,65,66].

**B. Geometric effect**

In the previous section, we see that polygonal metallic particles carry resonances that can be classified as edge modes, face modes, and hybrid modes. For a cylindrical particle, there are two plasmonic resonances: the classical dipole resonance (D mode), and the Bennett mode (M mode) [67-69]. It is interesting to see how the resonance modes in polygonal particles approach those of the cylinder as the number of corners increases. We will consider the effect of side number of the regular polygon while keeping either the radius $r_a$ fixed or the total number of electrons fixed. To see this geometric effect, we show the calculated absorption spectra for various N-sided polygons and cylindrical particles with the same radius in Fig. 6(a). The absorption spectra for these polygonal particles with the same total number of electrons are plotted in Fig. 6(b). We see that the edge and face modes come together and become the dipole mode as the side number $N$ increases. We note that the dipole resonance frequency of cylindrical particle is 3.96eV according to our calculations. This is consistent with Refs. [48,63], which is red-shifted comparing with classical resonance frequency $\omega_p / \sqrt{2} = 4.16$ eV (see Supplemental Material, Sec. VI [54]).

The Bennett mode of cylindrical particles corresponds to a perpendicular oscillation of the electron density in and out of the surface of the plasmonic particle along the radial direction. The induced charges have opposite signs but the same amplitude normal to the surface of one single face so that a single face is locally charge neutral for the Bennett mode, while that for the dipole mode is non-neutral [67-69]. The Bennett mode can only exist in models which allow the electron spill out effect (see Supplemental Material, Sec. VI [54]) so that charge can oscillate from inside the metal into the vacuum. According to our model, the resonance frequency of Bennett mode is 4.76eV, as shown by magenta lines in Fig. 6(a). From Figs. 6(a) and 6(b), we see that the hybrid modes of regular polygons can be treated as the precursors of the Bennett mode in the circular cylinder. The hybrid mode is very weak for the triangular particle, as shown by gray lines in Fig. 6(a). The hybrid mode absorption peak increases as side number of the polygon increases, as shown by orange, red, blue, and green lines in Fig. 6(a), eventually merging with the Bennett mode peak.

The results so far indicate that the edges in polygonal particles can strongly modify the plasmonic resonances of small particles, splitting the dipole resonance into edge/face modes that can be significantly red/blue shifted, depending on the angle of the wedge. Comparison of our SC-HDM results with classical model results (e.g. Fig. 3) also indicate that non-local effects are significant and cannot be ignored.

In addition to shape, size is obviously another parameter that can control plasmonic resonances. The size dependence of plasmonic modes directly relates to the dispersion relations of these plasmonic modes [70-72]. We plot the resonance frequencies of the $Ed_1^T$ mode and $Ed_1^S$ mode as a function of particle size $r_a^{-1}$ in Fig. 7(a) by solid red and blue lines. We see that the edge mode frequencies of both the triangle and square are nearly independent of the particle size in this range. This is consistent with the fact that the induced charges are mainly bounded to the single edge, so the size effect is minimal here. For comparison, we also calculate the size dependence of the $Ed_1^T$ mode and $Ed_1^S$ mode using the classical model, as shown by the dashed red and blue lines in Fig. 7(a). Similar to the SC-HDM, classical models also show little size dependence of these edge modes. However, the edge mode frequencies within SC-HDM are blue-shifted comparing to LRA. The physical reason is due to the location of the induced charge densities [73,74], which can be discussed using the Feibelman parameter that describes the centroid of the induced charges [63,73-74]. If the induced charges are mainly inside the jellium background, the resonance frequency blue-shifts. Conversely, if the induced charge mainly locates outside the jellium background, then the resonance frequency red-shifts. Figure 4 shows that the induced charge densities of the edge modes shrink into the jellium background, causing the blue-shift of those edge modes in SC-HDM.

We also plot the resonance frequencies of $Fa_1^T$ mode, $Fa_1^S$ mode, and dipole mode as a function of particle size $r_a^{-1}$ in Fig. 7(b) by red, blue, and green lines, respectively. In contrast to the edge modes, the face modes are red-shifted with the decreasing of particle size for both the triangle and square. The size dependence of the face modes is found to be similar to that of the dipole mode of cylindrical particles. It can be shown that the quantum corrections to the dipole mode of cylindrical particles are [47,63] (see also Supplemental Material, Sec. VII, for detailed derivations [54])

$$\left(\frac{\omega}{\omega_p}\right)^2 = \frac{1}{2}\left[1 - \frac{1}{r_a}(d_\perp - d_{//})\right], \qquad (4.1)$$

where $\omega_p^2 = e^2 n_{ion} / m_e \varepsilon_0$ is the bulk plasma frequency. $d_\perp$ and $d_{//}$ are the Feibelman parameters defined as

$$d_\perp = \frac{\int x \rho_1 dx}{\int \rho_1 dx}, \qquad d_{//} = \int \frac{n_0 - n_j}{n_{ion}} dx, \qquad (4.2)$$

for a single interface [73,74]. For charge neutral interfaces, $d_{//}$ is equal to zero, and $d_\perp$ describes the centroid of the charges induced by the external field. The calculated values of $d_\perp - d_{//}$ is shown in Fig. 7(e) for a sodium neutral flat interface. The interface is at $x = 0\text{nm}$, which is formed by sodium ($x < 0\text{nm}$) and vacuum ($x > 0\text{nm}$). The parameters used in this single interface calculations are the same with all the 2D calculations. It can be seen that at the dipole resonance frequency ($\sim 4.0\text{eV}$), $d_\perp > 0$, indicating that the resonance frequency of the dipole mode is red-shifted [63,73]. Our calculations also numerically verify this point, as shown by the solid green lines in Fig. 7(b). The induced charge densities shown in Fig. 5 of these face modes also mainly reside outside the jellium background, similar to the D mode of cylindrical particles. So the resonance frequencies of the face modes are red-shifted within SC-HDM. Furthermore, we also plot the size dependence of the Bennett mode in Fig. 7(c). It shows that the size dependence of Bennett mode is opposite to the dipole mode, consistent with Ref. [48].

To further investigate the edge modes, we plot the resonance frequencies of the lowest frequency edge modes calculated by SC-HDM as a function $\alpha$ by red stars in Fig. 7(d). For cylindrical particles ($\alpha = 90^o$), we plot the resonance frequency of dipole modes. We see that the edge mode frequencies increase monotonically with the interior angles $\alpha$. To see the physical reasons, we first recall the analytical dispersions of the even corner modes in the static and small wavenumber approximations are [61,62] (see also Supplemental Material, Sec. VIII, for detailed derivations [54])

$$\left(\frac{\omega}{\omega_p}\right)^2 = \frac{\alpha}{\pi}. \qquad (4.3)$$

We plot Eq. (4.3) as a solid line in Fig. 7(d). The qualitative agreement between analytical models and SCHDM results indicates the resonances in the spectrum are indeed edge modes. The quantitative deviations between these two are due to quantum corrections caused by electron spill out effect. To quantitatively model this effect, we employ the Feibelman *d-parameter* model based on the boundary element method (BEM) as defined in Ref. [63] to see the quantum corrections of these edge modes. The numerical results of this quantum BEM are shown by open blue circles in Fig. 7(d). We see excellent agreement between quantum BEM results and SC-HDM results for cylindrical particles. As the wedge angle decreases, both methods show a similar trend of deviation from the classical quasi-statics results, indicating the effect of electron spill out. However, the deviation between quantum BEM and SC-HDM results becomes larger for sharper corners. This is because local charging effect causes the $d_{//}$ to become polar angle dependent, which is inherent in SC-HDM but not included in the quantum BEM model. Nevertheless, quantum BEM still gives the correct shift directions compared with the analytical models. Fig. 7(d) reinforces the notion that the edge mode frequencies are very sensitive to the corner angles.

## C. Charged particles

The self-consistent hydrodynamics model can also be used to treat charged metallic particles. This is a distinct advantage of this method as there is no obvious way to consider charged particle within the framework of macroscopic classical electrodynamics. This is also useful as nanosized metallic particles are not necessarily neutral. To study charged particles, we need to carry out the calculations by changing the $\sigma$ manually in the constraint condition (2.7) for ground state calculations. For the sake of computation efficiency, we transform the constraint to the Neumann boundary condition of Eq. (2.5) (see Supplemental Material, Sec. II [54]). During the transformation, we assume that the electrostatic potential $\phi_0$ decays isotropically on the boundary $\partial\Omega_G$. This approximation is good when $r_g$ is chosen to be large enough and we choose the value of $r_g$ to make sure that further increase of $r_g$ does not change the absorption spectrum. The numerical results suggest that it is good to set $r_g$ to be 8.0 nm for 2nm particles (see Supplemental Material, Sec. IX [54]).

To see the effect of the additional charges on the plasmonic modes, we calculate the absorption cross sections of different charged cylinders, triangles, and squares, as plotted in Figs. 8(a), 8(b), and 8(c), respectively. The incident light is a single plane wave with incident angle $\theta_i = 0$. To make the additional charges $\sigma$ easier to interpret, we introduce the quantity $\Delta S$ defined as $\sigma/en_{\text{ion}}$ with the dimension of area in 2D configurations here. $\Delta S > 0$ ($< 0$) stands for positive (negative) charged particles. The ratio of $\Delta S$ to the total jellium area means the percentage of additional number of charge carriers to the total number of charge carriers. Figure 8 shows that all the plasmonic modes are blue-shifted in the positive charged particles, and red-shifted in the negative charged particles. However, the magnitudes of shifting are different for different modes. For the modes with long range charge oscillations, such as the D mode and face modes, the shifting is small compared to the "local" modes, such as Bennett mode and edge modes. This is because the additional charges accumulate near the surface and hence create a stronger effect on the surface than the bulk. Furthermore, the changes of the strength of Bennett mode are most significant in all these plasmonic modes. Since Bennett mode is locally charge neutral, so the doping should have the most prominent effects on this mode.

To illustrate the physics, we calculate and compare the $\text{Re}[d_\perp - d_{//}]$ of a single interface for charge neutral, positive charged, and negative charged cases in Fig. 7(e) by red, blue, and green lines, respectively. We see that near the dipole resonance frequency, $\text{Re}[d_\perp - d_{//}]$ is smaller in the positive charged interface than the neutral interface, so according to Eq. (4.1), the dipole plasmonic mode should blue-shift upon positive charging. For the negative charged interface, $\text{Re}[d_\perp - d_{//}]$ is bigger than the neutral interface around 4 eV and should induce a red shift. These effects explain the trend for the dipole mode in Fig. 8(a). The Bennett mode corresponds to the zero of $\text{Re}[d_\perp - d_{//}]$ because of its charge neutral nature. So the resonance frequency of Bennett mode should blue-shift for the positive charged particle, as seen from both Figs. 7(e) and 8(a).

**D. Energy dissipation distributions**

We have studied the plasmonic modes of polygonal particles by calculating absorption spectrum by integrating the fields at the far fields. Within the context of the hydrodynamic model, the energy absorption from the incident light is due to the work done by the EM wave on the electrons. In order to understand the energy dissipation distributions inside the electron gas, we calculate the time averaged work done $W_e = \langle \mathbf{J}_1 \cdot \mathbf{E}_1 \rangle$ by the fields on the electron gas. In our model, this work done could be obtained by the microscopic current $\mathbf{J}_1$ and the microscopic field $\mathbf{E}_1$. As a consistency check, the integration of $W_e / I_0$ ($I_0$ is the intensity of incident light) within the domain $\Omega_D$ should be equal to the absorption cross section calculated with fields at the far field, and we numerically verify this is indeed the case in Sec. X in Supplemental Material [54]. To see the distributions work done $W_e$ for different plasmonic modes, we plot in Figs. 9(a) to 9(f) the spatial distributions of $W_e/I_0$ for the plasmonic resonance modes shown in Fig. 6(a), including the dipole mode, Bennett mode, $\text{Ed}_1^T$ mode, $\text{Fa}_1^T$ mode, $\text{Ed}_1^S$ mode, and $\text{Fa}_1^S$ mode. From Fig. 9, we see the dominant loss of all these modes comes from the interface area of the particles, but not from the bulk interior areas. The difference between the modes is similar to the induced electron distributions, namely the energy dissipations mainly come from the edge (face) area for the edge (face) modes. From Figs. 9(d) and 9(f), we find standing wave like patterns in the energy dissipation distributions, which further indicate the size dependence of face modes is larger than that of edge modes.

**Section V. Conclusions**

In conclusion, we found three types of plasmonic modes for polygonal metallic particles, including edge modes, face modes and hybrid modes. These plasmonic modes are classified according to their induced electron concentration profile. We show that the edge modes are less sensitive to particle size than the face modes, but are more sensitive to corner angles of the edges. The edge and face modes can be viewed as broken-symmetry consequences of the dipole mode of a cylindrical particle. With the number of the sides of polygon increases, the edge mode move up and face mode move down in frequency to merge into the classical dipole plasmonic mode of a cylinder. The hybrid modes are found to be the precursor of the quantum Bennett mode. Finally, it is shown that positive (negative) charging causes the blue (red) shift to all these plasmonic modes.

**Acknowledgements**

This work is supported by Hong Kong Research Grants Council (grant no. AoE/P-02/12).

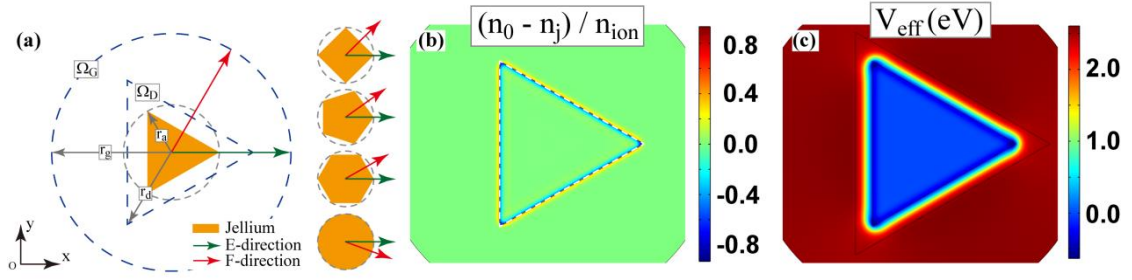

FIG. 1. (Color online) (a) Schematic picture of the plasmonic system under consideration, in which yellow region stands for the jellium background with circumradius $r_a$, $\Omega_G$ is the ground state calculation domain with radius $r_g$, and $\Omega_D$ is the excited state calculation domain with radius $r_d$. The right inset shows different regular polygons with the same circumradius $r_a$. The green (red) arrow denotes the edge (face) direction of the regular polygons. Calculated electron distribution $(n_0 - n_j)/n_{ion}$ and effective potential $V_{eff}$ of the ground state for a sodium triangle with radius $r_a = 2\,\text{nm}$ is shown in (b) and (c), respectively. The white dashed line in (b) highlights the jellium boundaries. The VW parameter $\lambda_\omega$ (see text) is chosen as 0.12, the radius of excited state calculation domain is $r_d = 2.7\,\text{nm}$, and that for ground state calculation domain is $r_g = 3.5\,\text{nm}$.

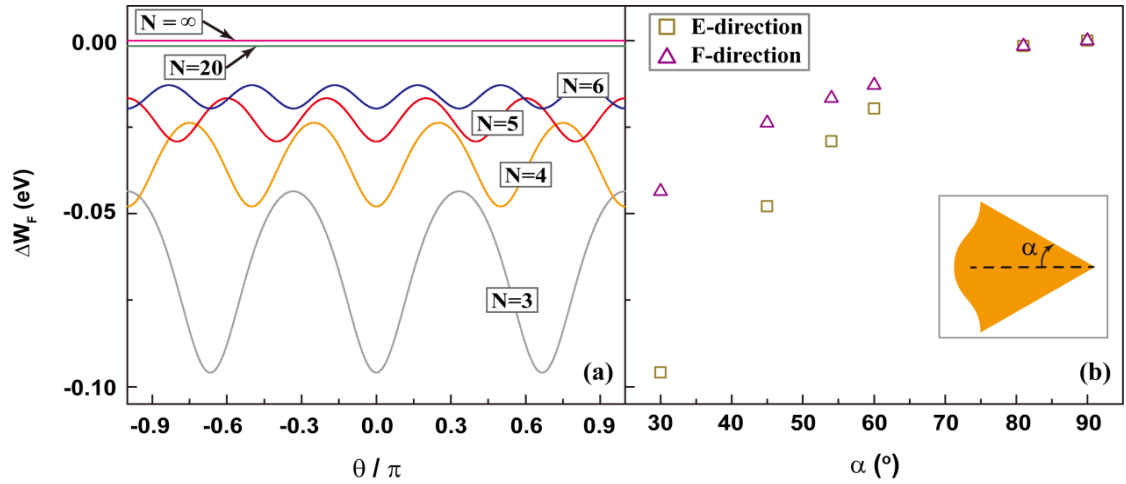

FIG. 2. (Color online) (a) Modulations of local work function as a function of polar angle $\theta/\pi$ for a triangle, square, pentagon, hexagon, 20-sided polygon, and cylinder are plotted by gray, orange, red, blue, green and magenta lines, respectively. (b) Local work functions in the E(edge)-direction and F(face)-direction as a function of interior angles $\alpha$ are plotted by open squares and open triangles, respectively. The circumradii of all the regular polygons are 2.0nm.

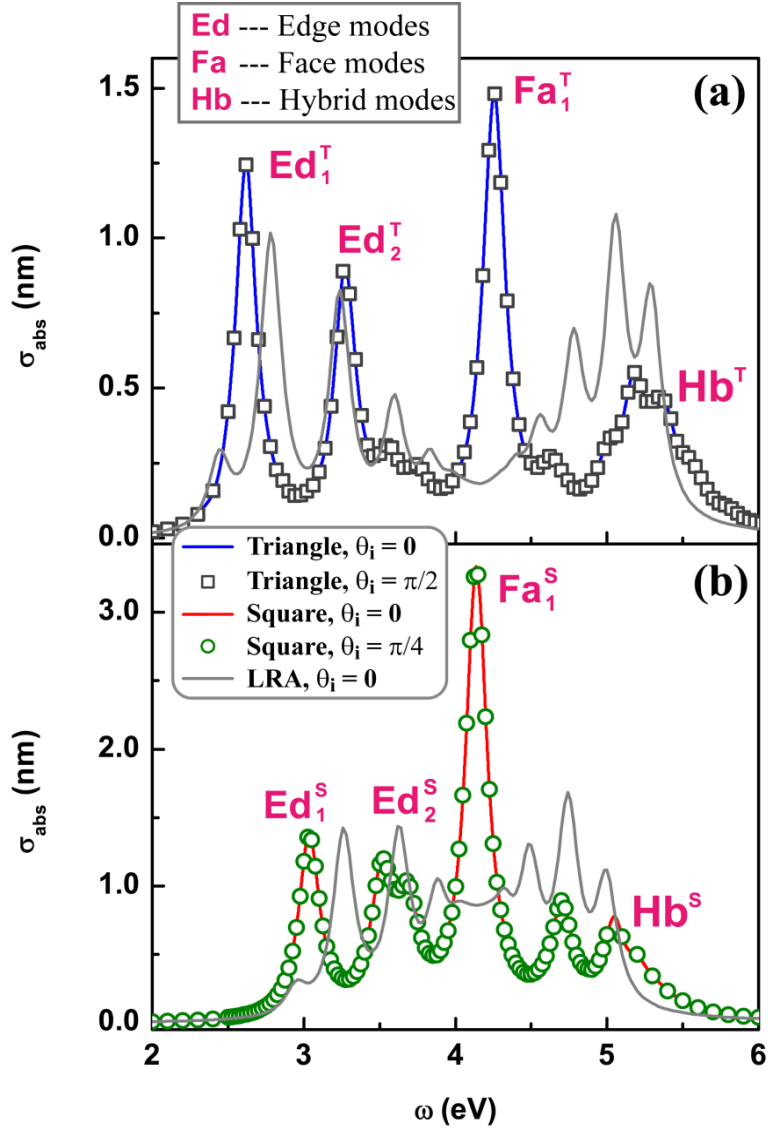

FIG. 3. (Color online) Absorption spectra of (a) a triangle and (b) a square with cicrcumradius $r_a = 2$ nm under plane wave incidence with incident angle $\theta_i = 0$ are plotted by blue and red solid lines, respectively. The loss parameter $\gamma$ is 0.17 eV throughout the domain $\Omega_D$. The absorption spectrum of triangle for plane wave with incident angle $\theta_i = \pi/2$ is plotted in (a) by gray dots, and that of square for plane wave with incident angle $\theta_i = \pi/4$ is plotted in (b) by green dots. The gray lines show the absorption spectra calculated by classical model. The resonance modes are labeled according to their induced charge profiles (see text). The superscript denotes the shapes (T stands for triangle) and the subscript denotes the mode number.

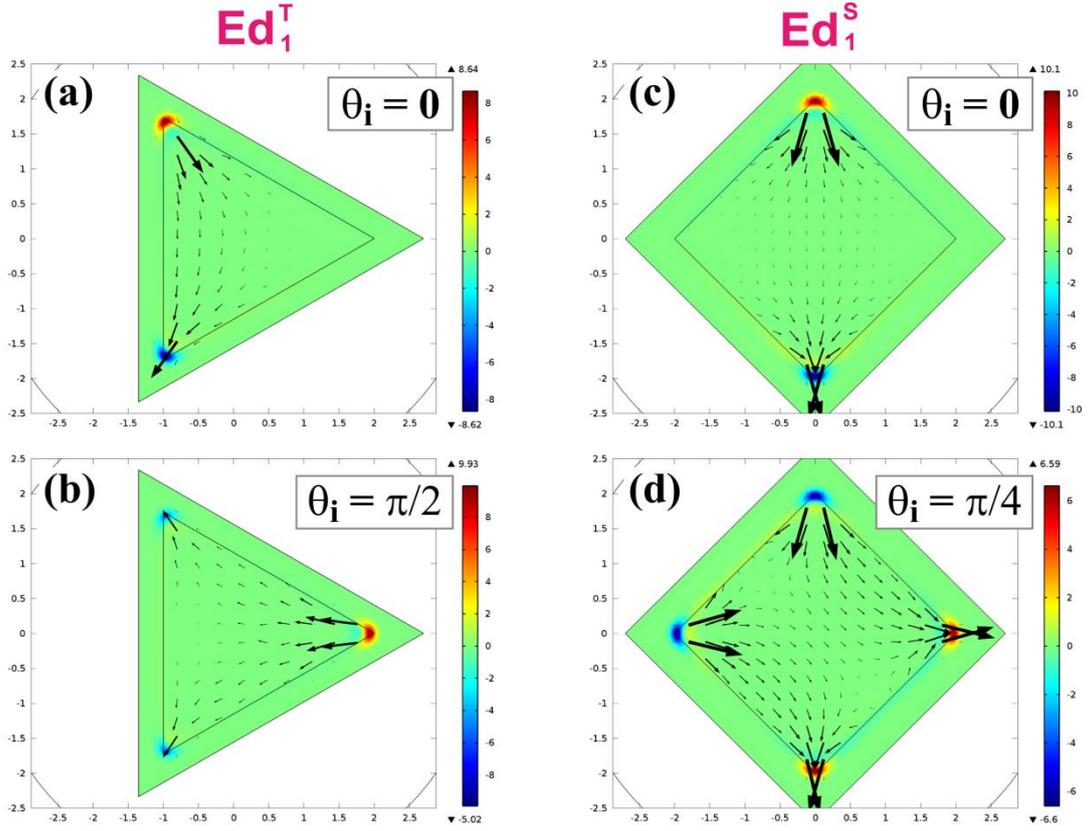

FIG. 4. (Color online) Plot of the induced electron density $\rho_1/E_0$ (color/intensity) and the current density $\mathbf{J}_1$ (vector field) at some particular time for the $Ed_1^T$ mode shown in Fig. 3 for a plane wave incident at an angle (a) 0 and (b) $\pi/2$, respectively. The induced electron density and current density for the $Ed_1^S$ mode are shown in (c) and (d) for plane wave incident angle 0 and $\pi/4$, respectively.

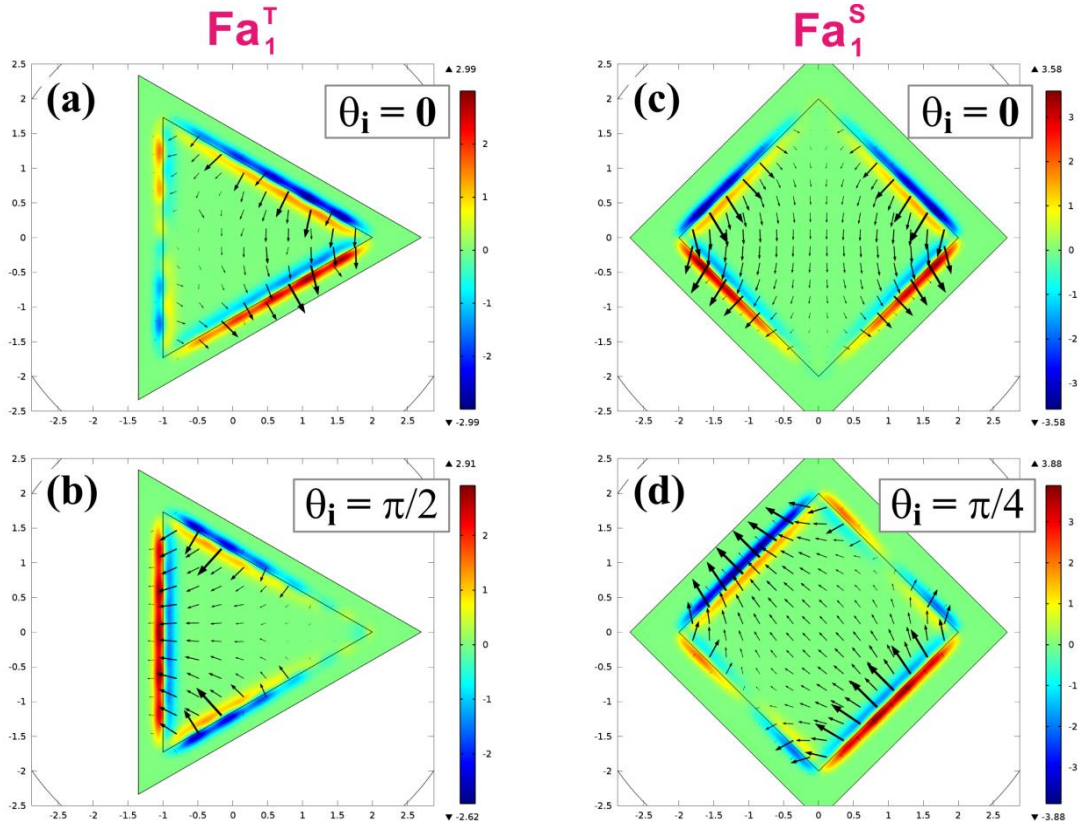

FIG. 5. (Color online) Plot of the induced electron density $\rho_1/E_0$ (color/intensity) and the current density $\mathbf{J}_1$ (vector field) at some particular time for the $Fa_1^T$ mode claimed in Fig. 3 are shown in (a) and (b) for plane wave incident at angles of 0 and $\pi/2$, respectively. The induced electron density and current density for the $Fa_1^S$ mode are shown in (c) and (d) for plane wave incident at angles 0 and $\pi/4$, respectively.

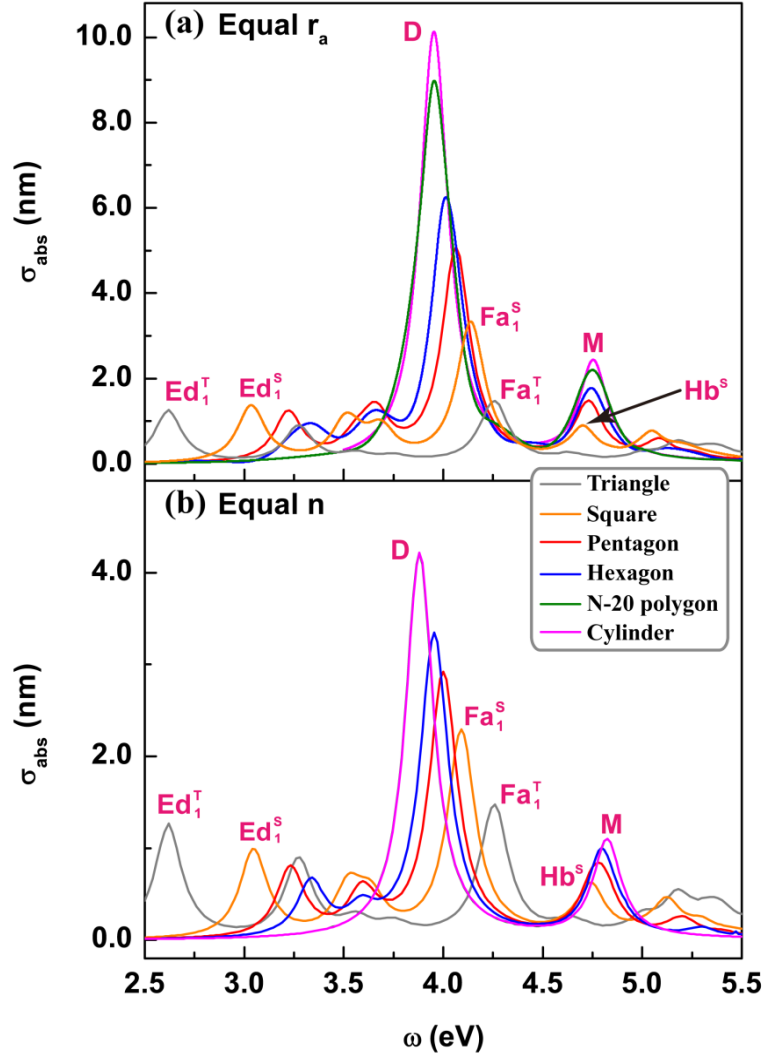

FIG. 6. (Color online) (a) Absorption spectra of regular polygons with increasing side number $N$ are plotted by the solid lines. All regular polygons have the same circumradius $r_a = 2.0$ nm, and loss parameter $\gamma$ is 0.17 eV throughout the domain $\Omega_D$. (b) Absorption spectra of polygons with increasing side number $N$ for fixed total number of electrons.

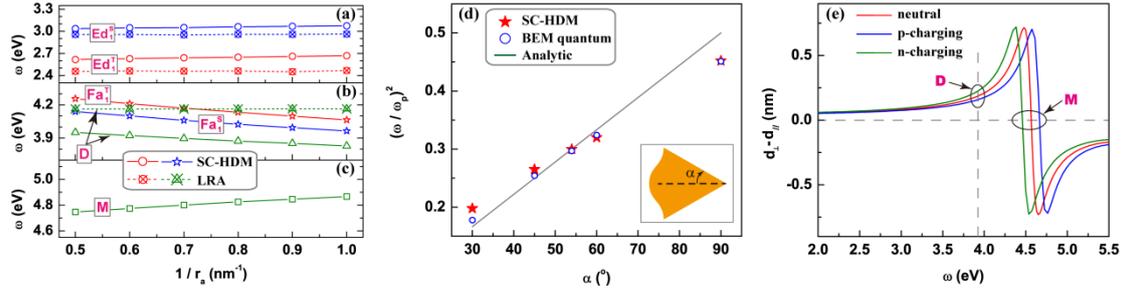

FIG. 7. (Color online) (a) The resonance frequencies of the edge modes as a function of the circumradius is plotted by red circles ($Ed_1^T$ mode) and blue circles ($Ed_1^S$ mode), respectively. The open circles are obtained by SC-HDM, and the circles with a cross inside are obtained using classical model (LRA). (b) The resonance frequencies of the face modes as a function of the radius are plotted by red stars ($Fa_1^T$ mode) and blue stars ($Fa_1^S$ mode), respectively. The resonance frequencies of the dipole mode (D mode) for cylindrical particles are also plotted by dark green triangles. (c) The resonance frequencies of the Bennett mode (M mode) for cylindrical particles are plotted by dark green squares. (d) The calculated resonance frequencies of the $Ed_1$ mode for SC-HDM as a function of interior angles $\alpha$ are plotted by solid stars. The solid line shows the analytical dispersions of edge modes for a single corner obtained under quasi-static and small wavenumber limitations. The blue circles show the numerical results obtained by BEM with quantum corrections under quasi-static limits. (e) The difference between $d_\perp$ and $d_{//}$ of a single interface for the charge neutral case, p-charging ($d_{//} = -0.001 \text{nm}$), and n-charging ($d_{//} = 0.001 \text{nm}$) are plotted by red, blue, and green lines, respectively.

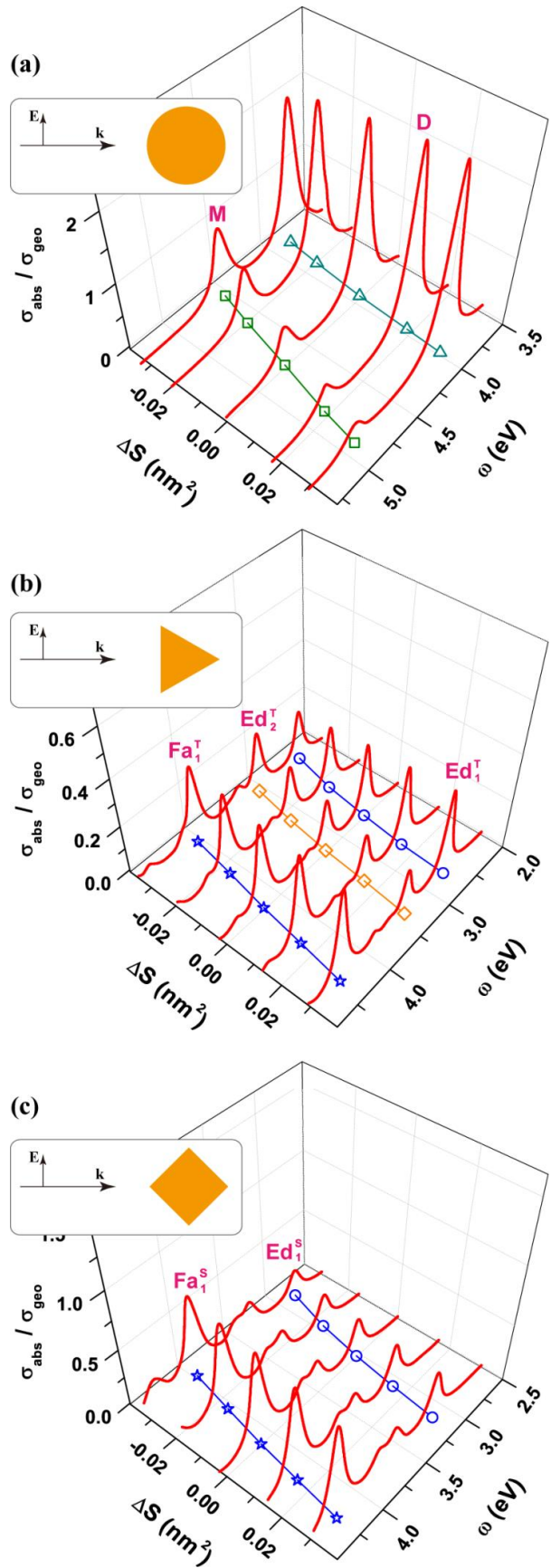

FIG. 8. (Color online) (a) Absorption spectra of a charged cylinder are shown by solid red lines for different additional amounts of charges. The open cyan triangle dots and open green square dots show the resonance frequencies as a function of $\Delta S$ for dipole mode (D) and Bennett mode (M), respectively. The vertical axis is the absorption cross section $\sigma_{abs}$ divided by geometric cross section $\sigma_{geo} = 2r_a$. (b) Absorption spectra of a charged triangle are shown by solid red lines for different additional amounts of charges. The dependence of resonance frequencies on $\Delta S$ are shown by open blue circles, open orange square dots and open blue star dots for the $Ed_1^T$ mode, the $Ed_2^T$ mode, and the $Fa_1^T$ mode, respectively. (c) Absorption spectra of charged square are shown by solid red lines for different additional amounts of charges. The dependence of resonance frequencies on the amount of $\Delta S$ are shown by open blue circles and open blue star dots for the $Ed_1^S$ mode and the $Fa_1^S$ mode, respectively. All the polygons have the same circumradius $r_a = 2.0$nm, and loss parameter $\gamma$ is 0.17 eV throughout the domain $\Omega_D$.

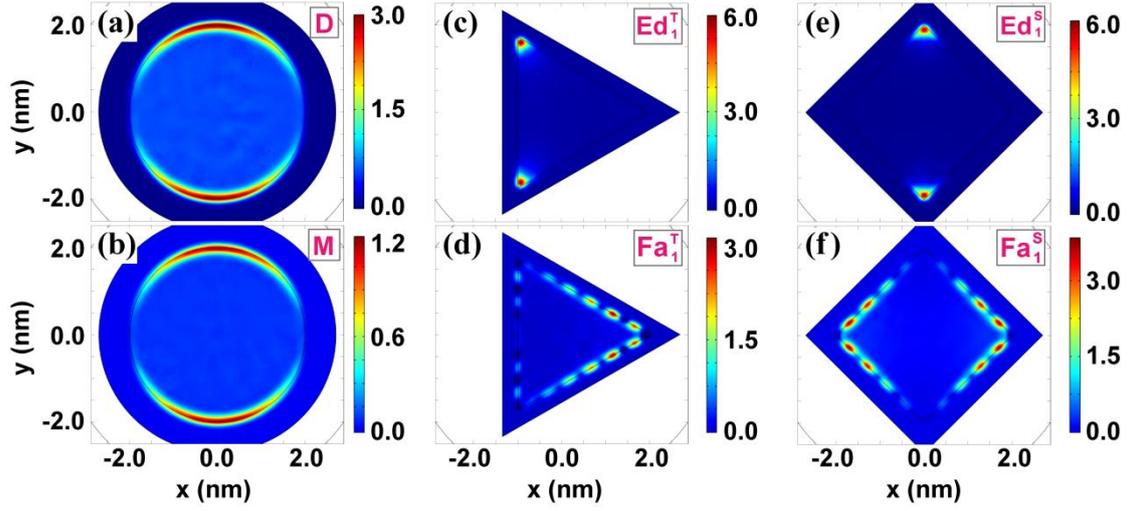

FIG. 9. (Color online) Contour plot of the energy dissipation $\langle \mathbf{J}_1 \cdot \mathbf{E}_1 \rangle / I_0$ in the unit of $\text{nm}^{-1}$ within the domain $\Omega_D$ for (a) Dipole mode (D), (b) Bennett mode (M), (c) $\text{Ed}_1^T$ mode, (d) $\text{Fa}_1^T$ mode, (e) $\text{Ed}_1^S$ mode, and (f) $\text{Fa}_1^S$ mode. These plasmonic modes corresponds to those in Fig. 6(a) with $r_a = 2.0 \text{nm}$.

# Supplemental Material --- Plasmonic modes of polygonal particles calculated using a quantum hydrodynamics method

Kun Ding and C. T. Chan[†]

*Department of Physics and Institute for Advanced Study,*
*The Hong Kong University of Science and Technology, Hong Kong*

† Corresponding E-mail: phchan@ust.hk

## Table of Contents

**Section I. Basic formulations of the self-consistent hydrodynamics model**

The systems studied are two dimensional (2D) polygonal particles, as shown in Fig. S1. The yellow region stands for the jellium background, in the shape of regular polygons. These polygons are bounded within by a circumscribed circle with radius $r_a$, as marked by gray dashed lines in Fig. S1. The dark-blue dashed cycle and triangle show the boundaries of ground state and scattering state calculation domain, respectively, denoted by $\Omega_G$ and $\Omega_D$.

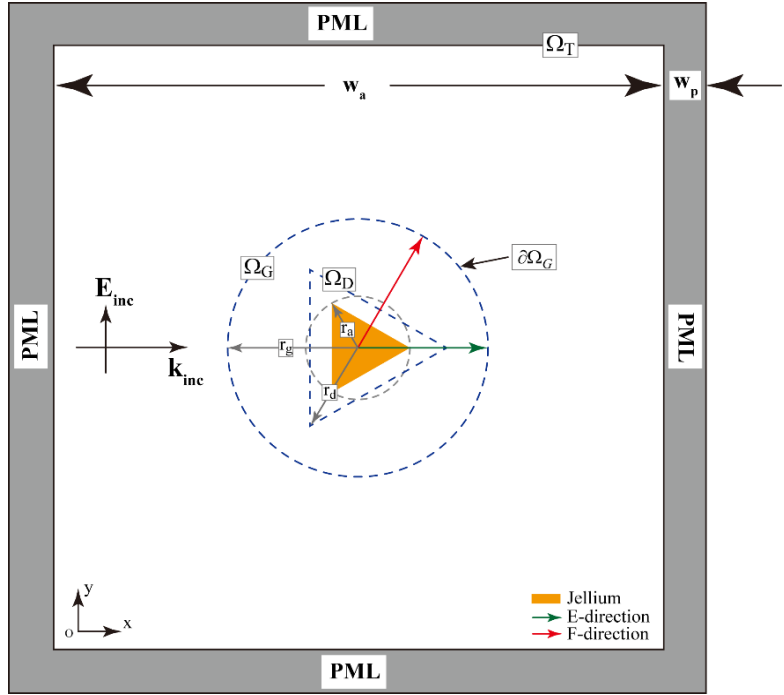

**Figure S1.** Schematic picture of the calculation domains and boundaries. The dashed lines stand for the virtual boundaries and the surrounding boundaries of the whole domain are perfect matched layer (PML).

We employ a self-consistent hydrodynamics model (SC-HDM) [1-5]. The basic assumption of hydrodynamics model (HDM) is to treat the electrons of metals as an ideal, irrotational, and isentropic fluid, which indicates that temperature effects are not included in the model. Under these assumptions, the electron gas can be fully described by the physical variables, electron density $n(\mathbf{r},t)$, and canonical momentum $\mathbf{p}(\mathbf{r},t)$. The equations of motion for the electron gas can be obtained from the time evolutions of these variables, namely [1-5],

$$m_e n\left(\frac{\partial \mathbf{v}}{\partial t} + \mathbf{v}\cdot\nabla\mathbf{v}\right) = -n\nabla\left(\frac{\delta G}{\delta n}\right) + nq_e(\mathbf{E} + \mathbf{v}\times\mathbf{B}), \quad (S1.1)$$

$$\frac{\partial n}{\partial t} + \nabla\cdot(n\mathbf{v}) = 0, \quad (S1.2)$$

where $q_e = -e$ is the charge of electron $(e > 0)$, $m_e$ is the mass of electron, $\mathbf{v} = (\mathbf{p} - q_e\mathbf{A})/m_e$ is the velocity of electron, $\mathbf{A}$ is the vector potential, and $\mathbf{E}, \mathbf{B}$ are the fields generated by the charges, currents and external electromagnetic waves. The energy functional $G[n(\mathbf{r},t)]$ plays the central role in the method. It contains kinetic energy, exchange energy, correlation energy terms. Here, we follow the choice in Ref. [2], but other forms of exchange and correlation energy can also be used in the method, such as Refs. [3,4].

The explicit form of the energy functional $G[n(\mathbf{r},t)]$ is taken as

$$G[n] = \int_{\Omega_G} g[n,\nabla n]\, \mathrm{d}\mathbf{r}, \quad (S1.3)$$

where

$$g[n,\nabla n] = \frac{3\hbar^2}{10m_e}(3\pi^2)^{2/3} n^{5/3} + \frac{\lambda_\omega \hbar^2}{8m_e}\frac{\nabla n\cdot\nabla n}{n} \\ -0.0588\frac{e^2 n^{4/3}}{\varepsilon_0} - \frac{0.035}{0.6024 + 7.8 a_\mathrm{H} n^{1/3}}\frac{e^2 n^{4/3}}{\varepsilon_0}. \quad (S1.4)$$

The first two terms are Thomas-Fermi (TF) and von Weizsacker (VW) kinetic energy functional, respectively [6]. TF term is the kinetic energy for a uniform system, while VW term is the leading correction of kinetic energy for the inhomogeneous electron gas. The third and fourth terms are Wigner's exchange and correlation energy functional [6,7]. By using the definition of functional derivative, the functional derivative of $G[n]$ is

$$\frac{\delta G}{\delta n} = \frac{\hbar^2}{2m_e}(3\pi^2 n)^{2/3} + \frac{\lambda_\omega \hbar^2}{4m_e}\left[\frac{\nabla n \cdot \nabla n}{2n^2} - \frac{\nabla^2 n}{n}\right]$$

$$-0.0588 \times \frac{4e^2}{3\varepsilon_0} n^{1/3} - \frac{4e^2}{3\varepsilon_0} \frac{0.035}{0.6024 + 7.8 a_H n^{1/3}} n^{1/3} \qquad (S1.5)$$

$$+\frac{7.8 a_H n^{2/3}}{3} \frac{0.035}{(0.6024 + 7.8 a_H n^{1/3})^2} \frac{e^2}{\varepsilon_0}.$$

Linear expansions of Eq. (S1.5) are needed to carry out ground state and scattering state calculations, and the explicit forms of these expansions will be given out later.

Equations (S1.1)-(S1.2) are the basic equations in the hydrodynamics model, and numerical solutions of them together with the Maxwell equations are needed to study the plasmonic behavior of our system. The next step is to do linear expansions of the following quantities in Eqs. (S1.1)-(S1.2) as

$$\begin{aligned} n &= n_0 + n_1, \qquad \mathbf{v} = \mathbf{v}_1, \\ \left(\frac{\delta G}{\delta n}\right) &= \left(\frac{\delta G}{\delta n}\right)_0 + \left(\frac{\delta G}{\delta n}\right)_1, \\ \mathbf{E} &= \mathbf{E}_0 + \mathbf{E}_1, \qquad \mathbf{B} = \mathbf{B}_1, \end{aligned} \qquad (S1.6)$$

where the subscript "0" stands for the equilibrium states and subscript "1" means the linear expansions of these quantities. The equilibrium $\mathbf{v}_0$ is zero which means no electrons move at equilibrium, and $\mathbf{B}_0$ is also zero because no external static magnetic field is imposed. The perturbation expansion is to substitute the expansions (S1.6) into Eqs. (S1.1) and (S1.2), and we arrive

$$m_e(n_0 + n_1)\left(\frac{\partial \mathbf{v}_1}{\partial t} + \mathbf{v}_1 \cdot \nabla \mathbf{v}_1\right)$$
$$= -(n_0 + n_1)\nabla\left[\left(\frac{\delta G}{\delta n}\right)_0 + \left(\frac{\delta G}{\delta n}\right)_1\right] + q_e(n_0 + n_1)(\mathbf{E}_0 + \mathbf{E}_1 + \mathbf{v}_1 \times \mathbf{B}_1), \qquad (S1.7)$$

$$\frac{\partial(n_0 + n_1)}{\partial t} = -\nabla \cdot \left[(n_0 + n_1)\mathbf{v}_1\right]. \qquad (S1.8)$$

Keeping the zero order terms in Eqs. (S1.7) and (S1.8), we arrive

$$-n_0 \nabla \left(\frac{\delta G}{\delta n}\right)_0 + q_e n_0 \mathbf{E}_0 = 0, \qquad (S1.9)$$

$$\partial_t n_0 = 0. \qquad (S1.10)$$

The static electric field $\mathbf{E}_0$ could be expressed as $\mathbf{E}_0 = -\nabla \phi_0$, which is produced by positive charges which contain ions and core electrons, and the s-band electrons considered here. The electrostatic potential $\phi_0$ obeys the following Poisson equation

$$\nabla^2 \phi_0 = \frac{q_e}{\varepsilon_0}(n_+ - n_0). \qquad (S1.11)$$

where $n_0$ is ground state electron density. The positive charge background $n_+$ is assumed to be jellium $n_+ = n_j$, namely $n_j = n_{\text{ion}}$ inside the metal, and $n_j = 0$ outside the metal. The ion density $n_{\text{ion}}$ is defined via dimensionless quantity $r_s$ as

$$n_{\text{ion}} = \frac{3}{4\pi (r_s a_{\text{H}})^3}, \qquad (S1.12)$$

where $a_{\text{H}} = 0.529\text{Å}$ is Bohr radius. Substituting $\mathbf{E}_0 = -\nabla \phi_0$ into Eq. (S1.9) could obtain

$$\left(\frac{\delta G}{\delta n}\right)_0 + q_e \phi_0 = \mu, \qquad (S1.13)$$

where $\mu$ is the chemical potential of the electron system. The total number of electrons imposes another constraint. Suppose the particle has a net charge $\sigma$, then the electron density should obey

$$\int_{\Omega_G} d\mathbf{r}\, e[n_+ - n_0] = \sigma. \qquad (S1.14)$$

Equations (S1.11), (S1.13) and (S1.14) collectively determine the ground state charge densities of the plasmonic particles.

The next step is to write out the first order terms in Eqs. (S1.7) and (S1.8) as the following

$$m_e n_0 \left(\frac{\partial \mathbf{v}_1}{\partial t}\right) = -n_0 \nabla \left(\frac{\delta G}{\delta n}\right)_1 + q_e n_0 \mathbf{E}_1, \qquad (S1.15)$$

$$\frac{\partial n_1}{\partial t} = -\nabla \cdot [n_0 \mathbf{v}_1]. \qquad (S1.16)$$

Denote $\mathbf{J}_1 = n_0 q_e \mathbf{v}_1$ and $\rho_1 = q_e n_1$, and then Eqs. (S1.15) and (S1.16) become

$$\frac{\partial \mathbf{J}_1}{\partial t} = \frac{e n_0}{m_e} \nabla \left( \frac{\delta G}{\delta n} \right)_1 + \frac{e^2 n_0}{m_e} \mathbf{E}_1, \tag{S1.17}$$

$$\frac{\partial \rho_1}{\partial t} + \nabla \cdot \mathbf{J}_1 = 0. \tag{S1.18}$$

Equation (S1.17) is the response equation, and Eq. (S1.18) is the continuity condition for induced charges. There are no explicit loss mechanisms in the method, so we introduce the loss phenomenally by adding the term $-\gamma \mathbf{J}_1$ to the right side of Eq. (S1.17). Normally, $\gamma$ is a constant only inside the metal domain, but now $\gamma$ should exist in all the space because electrons can spill out into the vacuum [8]. The consequence of this $\gamma$ is that dissipations exist in the space where the electron density is nonzero, namely $n_0 \neq 0$. A more accurate setting is that $\gamma$ depends on $n_0$, but we treat it as a constant in all the space here [9]. Furthermore, assume the time factor is $e^{-i\omega t}$, then Eqs. (S1.17) and (S1.18) become

$$(-i\omega + \gamma)\mathbf{J}_1 = \frac{e n_0}{m_e} \nabla \left( \frac{\delta G}{\delta n} \right)_1 + \frac{e^2 n_0}{m_e} \mathbf{E}_1, \tag{S1.19}$$

$$\nabla \cdot \mathbf{J}_1 - i\omega \rho_1 = 0, \tag{S1.20}$$

Maxwell equations coupled with these linear response equations are

$$\nabla \times (\nabla \times \mathbf{E}_1) - \left( \frac{\omega}{c} \right)^2 \mathbf{E}_1 = i\omega \mu_0 \mathbf{J}_1, \tag{S1.21}$$

where $\gamma$ is the loss parameter introduced empirically here, and the electric fields $\mathbf{E}_1 = \mathbf{E}_{inc} + \mathbf{E}_{sca}$ are total fields including incident fields and scattering fields from the environment. Equations (S1.19), (S1.20), and (S1.21) determine the linear response of the plasmonic particles.

Two steps are needed numerically to obtain the plasmonic resonances and related properties of the system using SC-HDM. We first solve Eqs. (S1.11), (S1.13), and (S1.14) to get the ground states, and then solve Eqs. (S1.19), (S1.20), and (S1.21) to get the scattering states. In the following sections, we will discuss the numerical implementations and some details of the model.

**Section II. Explicit forms and weak forms of ground state equations**

To obtain the ground state electron distributions, Eqs. (S1.11), (S1.13) and (S1.14) are solved numerically by the finite element method (FEM). To make the equation simpler, we introduce the constants $k_F$, $v_F$, $k_{TF}$, and $\omega_p$ as the following

$$E_F = \frac{\hbar^2 k_F^2}{2m_e} = \frac{1}{2}m_e v_F^2 = \frac{\hbar^2}{2m_e}(3\pi^2 n_{ion})^{2/3}, \tag{S2.1}$$

$$k_{TF} = \frac{\omega_p}{v_F} \qquad \omega_p^2 = \frac{e^2 n_{ion}}{m_e \varepsilon_0}. \tag{S2.2}$$

The quantities $k_F$, $v_F$, $k_{TF}$, and $\omega_p$ should be defined via electron density $n_0$, indicating they are position dependent variables, but we use ion density instead to make all these parameters as constants. Throughout this work, we will use this definition.

To make these equations in a more transparent form, we introduce some dimensionless quantities as

$$f_0 = \sqrt{\frac{n_0}{n_{ion}}} \qquad \phi_0' = \frac{\phi_0 \epsilon_0 k_{TF}^2}{e n_{ion}} \qquad \mu' = \frac{\mu}{E_F}. \tag{S2.3}$$

By multiplying $f_0$ to Eq. (S1.13), this equation becomes [2]

$$-\frac{\hbar^2}{2\lambda_\omega^{-1} m_e}\nabla^2 f_0 + V_{eff} f_0 = 0, \tag{S2.4}$$

where $\lambda_\omega^{-1} m_e$ plays the role of effective mass, the effective potential $V_{eff}$ is

$$V_{eff} = E_F \left[ f_0^{4/3} + C_0 f_0^{4/3} - C_1 f_0^{2/3} - 2\phi_0' - \mu' \right], \tag{S2.5}$$

and some coefficients are defined as

$$C_0 = \frac{2}{3}\frac{0.035 X_1}{(0.6024 + X_1 f_0^{2/3})^2}\frac{k_{TF}^2}{n_{ion}^{2/3}}, \tag{S2.6}$$

$$C_1 = \frac{8}{3}\left(0.0588 + \frac{0.035}{0.6024 + X_1 f_0^{2/3}}\right)\frac{k_{TF}^2}{n_{ion}^{2/3}},\qquad (S2.7)$$

$$C_2 = \frac{\lambda_\omega}{k_F^2} \qquad X_1 = 7.8 a_H n_{ion}^{1/3}. \qquad (S2.8)$$

Using the definitions above, the Poisson equation (S1.11) could be written as

$$\frac{1}{k_{TF}^2}\nabla^2 \phi_0' = f_0^2 - f_{ion}^2, \qquad (S2.9)$$

where $f_{ion} = 1$ in the metal domain, and $f_{ion} = 0$ in the vacuum. Furthermore, the constraint equation (S1.14) could be written as

$$\int_{\Omega_G} d\mathbf{r}\left[f_{ion}^2 - f_0^2\right] = \frac{\sigma}{en_{ion}} \equiv \Delta S, \qquad (S2.10)$$

where $\Omega_G$ is the ground state calculation domain, and $\Delta S$ has the same unit with volume in 3D.

The next step is to implement the equations (S2.4) and (S2.9) to weak forms used in FEM [10]. By multiplying test functions to the equations and integrating over the calculation domain $\Omega_G$, then the weak forms of Eqs. (S2.4) and (S2.9) are

$$\text{Wk}(f_0) = C_2 \nabla f_0 \cdot \nabla \tilde{f}_0 + \frac{V_{eff}}{E_F}\tilde{f}_0 f_0, \qquad (S2.11)$$

$$\text{Wk}(\phi_0') = \frac{1}{k_{TF}^2}\nabla \phi_0' \cdot \nabla \tilde{\phi}_0' + \tilde{\phi}_0'\left(f_0^2 - f_{ion}^2\right), \qquad (S2.12)$$

where $\text{Wk}(u)$ and $\tilde{u}$ denote the weak form and test function of an arbitrary function $u$. Then the next issue is how to implement the constraint Eq. (S2.10). The straightforward way is to treat the integration as Lagrange multiplier in the weak form, but it is not easy to get converge results by using this method. So we rewrite it as the boundary conditions of $\phi_0'$. By integrating Eq. (S2.9) in the domain $\Omega_G$, and using Stokes's theorem, the constraint equation (S2.10) becomes

$$\int_{\partial\Omega_G} d\mathbf{A} \cdot \frac{\nabla \phi_0'}{k_{TF}^2} = -\Delta S, \qquad (S2.13)$$

where the integration is performed on the surface $\partial\Omega_G$ enclosed the domain $\Omega_G$. Afterwards, we introduce an approximation, the gradient of electrostatic potential $\phi_0'$ is isotropic in all the directions on the boundary $\partial\Omega_G$. Based on this, Eq. (S2.13) becomes the boundary condition as [11]

$$-\hat{n}\cdot\frac{\nabla\phi_0'}{k_{TF}^2}=\frac{\Delta S}{A}, \quad (S2.14)$$

where $\hat{n}$ denotes the normal direction of the closed surface, and $A$ is the total area of the closed surface. In other words, Eq. (S2.14) is the Neumann boundary condition of the partial differential equation (PDE) (S2.9). For the PDE (S2.4), the boundary condition is the Dirichlet boundary condition, namely $f_0=0$.

In summary, numerical solution of ground state is to write weak forms Eqs. (S2.11) and (S2.12) in COMSOL [12] for the unknowns $(f_0,\phi_0')$, and apply the boundary conditions (S2.14) and $f_0=0$.

**Section III. Explicit forms and weak forms of scattering state equations**

To solve the scattering state equations (S1.19), (S1.20), and (S1.21), first we need to perform linear expansions on $n_1$ in $\left(\frac{\delta G}{\delta n}\right)_1$, and express the first term of the right side in Eq. (S1.19) as

$$\frac{en_0}{m_e}\nabla\left(\frac{\delta G}{\delta n}\right)_1=\frac{1}{3}ev_F^2\left[\nabla-\frac{2\nabla f_0}{f_0}\right]Q(n_1), \quad (S3.1)$$

where the new function $Q(n_1)$ is

$$Q(n_1)=D_{e1}n_1+\frac{3}{2}C_2\frac{\nabla f_0\cdot\nabla n_1}{f_0}-\frac{3}{4}C_2\nabla^2 n_1, \quad (S3.2)$$

and the coefficients are defined as

$$D_{e1}=f_0^{4/3}+\frac{3}{2}C_2\left[\frac{\nabla^2 f_0}{f_0}-\frac{\nabla f_0\cdot\nabla f_0}{f_0^2}\right]-\frac{1}{2}C_1f_0^{2/3}+3C_0f_0^{4/3}-C_3f_0^2, \quad (S3.3)$$

$$C_3 = \frac{2}{3} \frac{0.035 X_1^2}{(0.6024 + X_1 f_0^{2/3})^3} \frac{k_{TF}^2}{n_{ion}^{2/3}}. \tag{S3.4}$$

It should be mentioned that $Q(n_1)$ only contains the linear order terms of $n_1$, but Eq. (S3.1) contains third order derivative of $n_1$, so in order to solve them, we introduce an auxiliary quantity $\mathbf{F}$ and electric polarization $\mathbf{P}$ to replace $(n_1, \mathbf{J}_1)$ as [4]

$$\mathbf{F} = \nabla n_1 \qquad \mathbf{J}_1 = -i\omega \mathbf{P} \qquad n_1 = \frac{1}{e} \nabla \cdot \mathbf{P}. \tag{S3.5}$$

Based on all of these, Eqs. (S1.19), (S1.20), and (S1.21) could be rewritten as

$$\frac{1}{3} e v_F^2 \left[ \nabla - \frac{2\nabla f_0}{f_0} \right] Q(n_1) + \frac{e^2 n_0}{m_e} \mathbf{E}_1 + \omega(\omega + i\gamma)\mathbf{P} = 0, \tag{S3.6}$$

$$\nabla(\nabla \cdot \mathbf{P}) - e\mathbf{F} = 0, \tag{S3.7}$$

$$\nabla \times (\nabla \times \mathbf{E}_1) - \left(\frac{\omega}{c}\right)^2 \mathbf{E}_1 = \mu_0 \omega^2 \mathbf{P}, \tag{S3.8}$$

where the function $Q(n_1)$ is

$$Q(n_1) = \frac{D_{e1}}{e} \nabla \cdot \mathbf{P} + \frac{3}{2} C_2 \frac{\nabla f_0 \cdot \mathbf{F}}{f_0} - \frac{3}{4} C_2 \nabla \cdot \mathbf{F}. \tag{S3.9}$$

Similar to the derivations in section II, the weak forms of Eqs. (S3.6) and (S3.7) are

$$\text{Wk}(\mathbf{P}^j) = \tilde{\mathbf{P}}^j \left[ \varepsilon_0 \omega_p^2 f_0^2 \mathbf{E}_1^j + \omega(\omega + i\gamma)\mathbf{P}^j \right] \\ - \frac{1}{3} e v_F^2 \left[ (\partial_j \tilde{\mathbf{P}}^j) + \frac{2(\partial_j f_0)}{f_0} \tilde{\mathbf{P}}^j \right] Q(n_1), \tag{S3.10}$$

$$\text{Wk}(\mathbf{F}^j) = -(\partial_j \tilde{\mathbf{F}}^j)(\nabla \cdot \mathbf{P}) - e\tilde{\mathbf{F}}^j \mathbf{F}^j, \tag{S3.11}$$

where the superscript $j$ denotes the the $j$-th ($j = x, y, z$) component of some vector, the subscript $j$ denotes partial derivative to $x_j$, and Einstein summation rule is NOT applied here. For Eq. (S3.8), we could add the additional source term $\mathbf{P}$ to the existing EM module of COMSOL, so there is no need to rewrite weak form of Maxwell equations. It should be pointed out the time factor in COMSOL is $e^{i\omega t}$, so when writing the weak forms in COMSOL, one need to change the $i$ to $-i$.

As shown in Fig. S1, we divide the total calculation domain $\Omega_T$ into two subdomains, one is the calculation domain $\Omega_D$ within which we implement the weak forms (S3.10) and (S3.11) as the homemade module, and couple this module with the existing EM module by adding the source term $\mathbf{P}$. Another subdomain $\Omega_F \left( \equiv \Omega_T - \Omega_D \right)$ is the free space within which we only use existing EM module without any source terms. This kind of setting could avoid huge computational resource, but it is crucial to set the boundary conditions between these two domains. The boundary conditions of electromagnetic fields between these two domains are continuous, so it is no need to specify for Eq. (S3.8). For the induced charges, the physical constraint on the interface is $n_1 = 0$, so the boundary condition is $\hat{n} \cdot \mathbf{P} = 0$ for Eq. (S3.6). For Eq. (S3.7), no physical constraints exist because it is an auxiliary equation, so no boundary conditions are needed between these two domains. The outer boundary conditions of $\Omega_T$ are perfect matched layer (PML).

**Section IV. The von Weizsacker parameter $\lambda_\omega$**

The parameter $\lambda_\omega$ comes from von Weizsacker (VW) kinetic energy functional, and in the ground state equations (S2.4), it plays the role of effective mass. A smaller $\lambda_\omega$ indicates a larger effective mass of the s-band electron, leading to weaker spill out effect of electrons near the metal surface [6,13]. To verify this point, we calculate the ground state of electrons for a sodium slab, as shown in Fig. S2. From Fig. S2(a), we see that for the smaller $\lambda_\omega$, the electron tails in the vacuum is shorter, and also the charge density oscillation inside the jellium region are damped more quickly. This will also cause the static surface dipole to be weaker, leading to a lower work function for a smaller $\lambda_\omega$, as shown in Figs. S2(b) and S2(d). Furthermore, it is worthy to check the boundary conditions claimed in section II. So we integrate the $\left[ f_0^2 - f_{\text{ion}}^2 \right]$ in the calculation domain, as shown in Fig. S2(c). It can be seen that $d_\parallel$ equals to zero for all the $\lambda_\omega$, indicating that the slab is indeed charge neutral one.

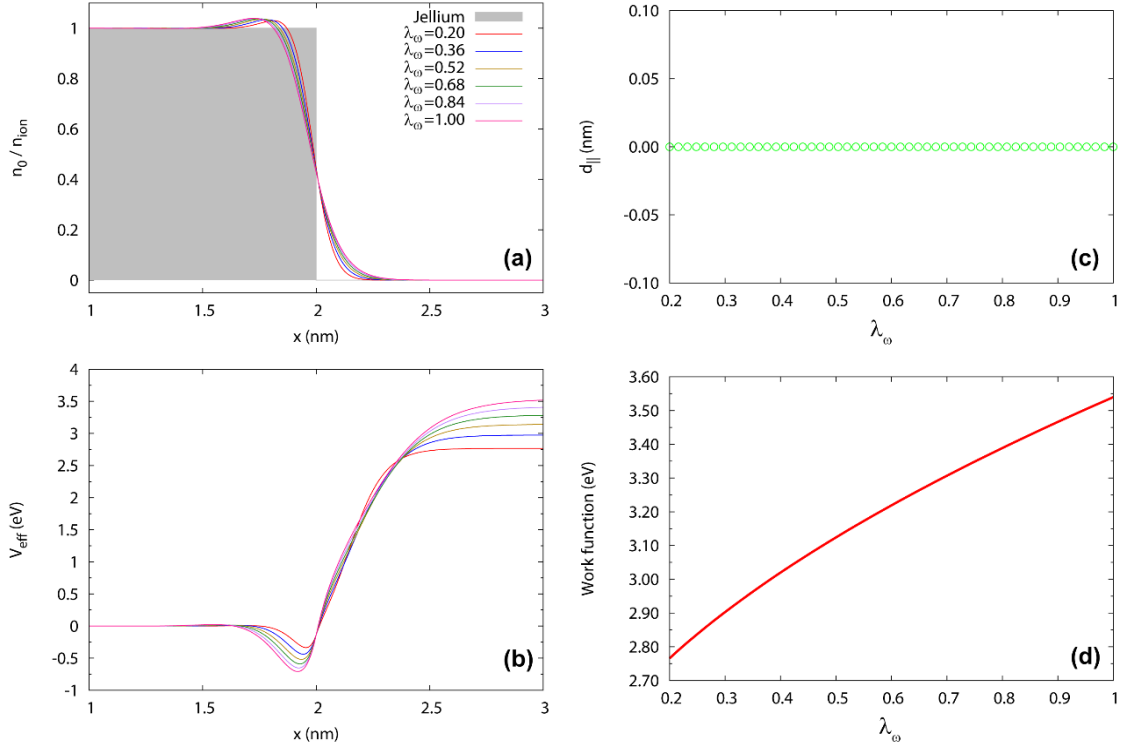

**Figure S2.** Electron distributions $n_0/n_{ion}$ and effective potential $V_{eff}$ for a sodium slab are plotted in (a) and (b), respectively, for different values of VW parameters $\lambda_\omega$. The thickness of the slab is 4.0 nm and its center is at x=0.0 nm. (c) The d-parameters $d_\parallel$, defined as $d_\parallel \equiv \int dx \left[ f_0^2 - f_{ion}^2 \right]$, as a function of $\lambda_\omega$ is plotted by open green circles. (d) Work function as a function of $\lambda_\omega$ is shown by red line.

To study the effect of $\lambda_\omega$ on the plasmonic modes, we also calculate absorption cross sections of a sodium cylindrical particle under plane wave illumination ($\theta_i = 0$) for different values of $\lambda_\omega$, as shown in Fig. S3. It can be seen that the common behavior of these curves is the same: they all have two peaks. The lower one is the dipole mode, the higher one is the Bennett mode. For decreasing values of $\lambda_\omega$, both dipole peak and Bennett peak are blue-shifted, and the dipole peak goes stronger, and the Bennett peak weaker [9].

The most straightforward way to set the value of $\lambda_\omega$ is to compare these results with ab-initio calculations. If we want to make the work function correct, $\lambda_\omega$ should set to 0.435. However, setting $\lambda_\omega$ to a smaller value of 1/9 would give a better agreement with the absorption spectrum measured experimentally. Several prior papers attempted to solve this well-known dilemma, but there are still problems [1-5]. We believe that this is an intrinsic problem of this model due to the fact that effective masses cannot be a constant for all the electrons. In our calculations, we choose $\lambda_\omega$ to 0.12 because further decreasing $\lambda_\omega$ will cause the convergence of ground state calculations for complicated particles more difficult, but does NOT change the absorption spectrum very much, as shown in Fig. S3.

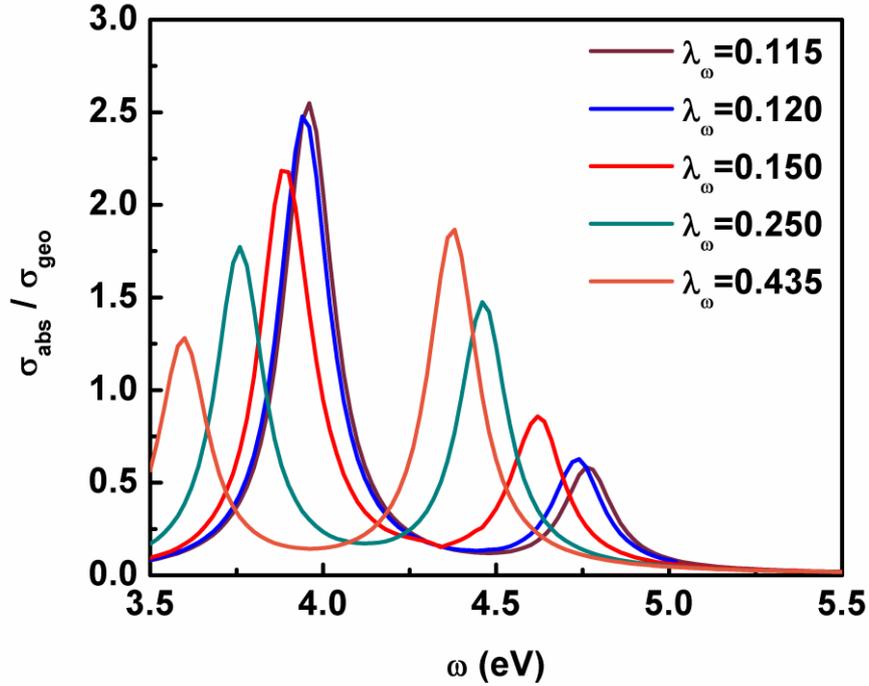

**Figure S3.** Absorption spectra of a sodium cylinder for different VW parameter $\lambda_\omega$ are shown by solid lines. The radius of cylinder is 2.0 nm, and loss parameter $\gamma$ is 0.17 eV throughout the domain $\Omega_D$. The incident light is a plane wave with incident angle $\theta_i = 0$.

**Section V. Discussion of the scattering calculation domain $\Omega_D$**

In this section, we discuss the setting of the scattering calculation domain $\Omega_D$. It should be pointed out that the calculation of $Q(n_1)$ and $D_{e1}$ involves a challenging numerical problem, namely the evaluation of $\nabla f_0 / f_0$ in the vacuum region. This is a "division by zero" type problem because the electron distribution decays exponentially in the vacuum. However, the value of $\nabla f_0 / f_0$ should be a finite meaningful number because the number describes the decay rate of the ground state electrons in vacuum, which is proportional to the work function. So before doing the scattering state calculations, we plot the distributions of $D_{e1}$ in the domain $\Omega_G$ in Fig. S4(a) for a charge neutral triangle. It can be seen that the values are still well defined near the particle except near the boundaries of the domain $\Omega_G$. By recognizing this, we choice the domain $\Omega_D$ to make sure that the value of $\nabla f_0 / f_0$ is still well defined within it. To verify this numerically, we plot absorption spectrum for a charge neutral triangle with different size of the domain $\Omega_D$ by blue and red lines in Fig. S4(b). We see that the absorption spectra are similar, especially the resonance peaks. So we set the radius $r_d$ of $\Omega_D$ to $r_d = r_a + 0.7\text{nm}$ throughout this work.

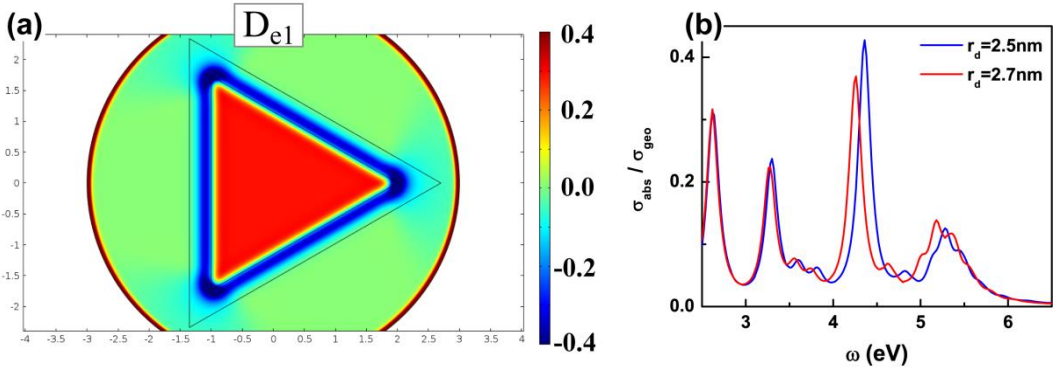

**Figure S4.** (a) Color plot of the coefficient $D_{e1}$ in the domain $\Omega_G$ for a charge neutral triangle with circumradius 2.0 nm. (b) The blue and red lines plot the absorption spectra of the triangle for $r_d = 2.5\text{nm}$ and $r_d = 2.7\text{nm}$, respectively. The incident light is a plane wave with incident angle $\theta_i = 0$.

**Section VI. Comparison of different models for a cylindrical particle**

In this section, we will compare our model used in the main text to other known models, including classical model and hard-wall hydrodynamics model (HW-HDM) [14-17]. The classical model means that we ignore the functional $G[n]$ in Eq. (S1.19), Eq. (S1.19) then gives a local conductivity as

$$\sigma(\omega) = \frac{\mathbf{J}_1}{\mathbf{E}_1} = \frac{i}{\omega + i\gamma} \frac{e^2 n_0}{m_e}. \tag{S6.1}$$

Till now, the value of $n_0$ is undefined because we do not know the ground states. The next assumption is to set $n_0$ equals the jellium density, meaning that no spill-out effect is allowed here. By using this, the local conductivity of Eq. (S6.1) becomes zero in vacuum, and becomes Drude model inside the metal, namely

$$\sigma(\omega) = \frac{i}{\omega + i\gamma} \frac{e^2 n_{\text{ion}}}{m_e} = \frac{i\omega_p^2 \varepsilon_0}{\omega + i\gamma} \Rightarrow \varepsilon(\omega) = 1 + i\frac{\sigma}{\omega \varepsilon_0} = 1 - \frac{\omega_p^2}{\omega(\omega + i\gamma)}. \tag{S6.2}$$

This gives the classical calculation theme. Since the model assumes that conductivities are determined by local currents and electric fields, so this assumption is called as local response approximation (LRA).

The so-called hard-wall hydrodynamics model goes beyond classical model by employing the Thomas-Fermi kinetic energy in $G[n]$, but $n_0$ is still set to the jellium density, which means the surface potential barrier is infinite and there is no spill out effect. This is why this model is called as hard-wall hydrodynamics model [14-17]. Based on these two points, Eq. (S1.19) becomes

$$(-i\omega + \gamma)\mathbf{J}_1 = \varepsilon_0 \omega_p^2 \mathbf{E}_1 - \beta^2 \nabla \rho_1, \tag{S6.3}$$

where $\beta^2 = (3/5)v_F^2$. It should be pointed out that the multiplication of a factor 9/5 to the energy functional $G[n]$ is used in Eq. (S6.3). This is to correct the high-frequency parts of this model [18]. Substituting continuity equations into Eq. (S6.3), we arrive

$$\omega(\omega + i\gamma)\mathbf{J}_1 + \beta^2 \nabla(\nabla \cdot \mathbf{J}_1) - i\omega \varepsilon_0 \omega_p^2 \mathbf{E}_1 = 0. \tag{S6.4}$$

This is the response equation to be solved in HW-HDM. There still exist an additional boundary condition, namely $\hat{n} \cdot \mathbf{J}_1 = 0$ on the metal surface because the electrons cannot go outside the metal region.

To compare these models, we plot the absorption spectrum of a charge neutral cylinder under plane wave illumination ($\theta_i = 0$) in Fig. S5. The red, green, and blue lines are calculated by using LRA, HW-HDM, and SC-HDM, respectively. The resonance peak of LRA is at $\sim \omega_p / \sqrt{2}$, and the peaks of HW-HDM and SC-HDM are shifted in opposite directions. The most important difference of these models is that the Bennett mode is absence in both LRA and HW-HDM. This is consistent with the fact that no electron spill out is allowed in either LRA or HW-HDM, and hence an important mechanism is missing. Another issue is the loss in the SC-HDM. We could assume the loss parameter $\gamma$ only exist in the jellium background or throughout the domain $\Omega_D$. We plot the absorption spectrum for these two setting in Fig. S5 by magenta and blue lines. It can be seen that these two setting do NOT affect the resonance peaks, so throughout this work, we will assume the loss parameter $\gamma$ exist throughout the domain $\Omega_D$.

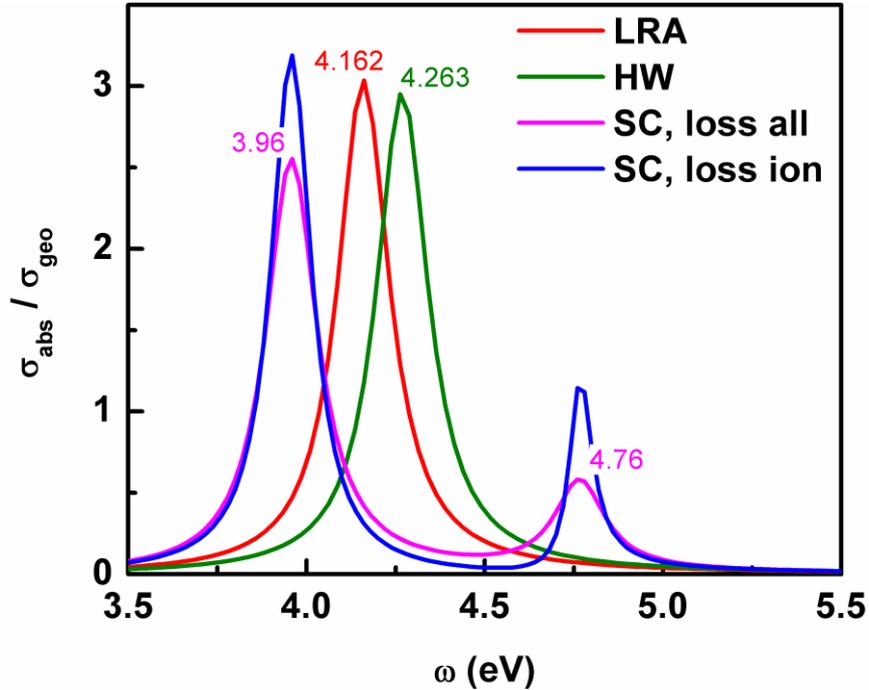

**Figure S5.** Absorption spectra of a sodium cylinder calculated by classical model (LRA) and hard-wall hydrodynamics model are shown by red line and green line, respectively. The radius of the cylinder is
16

2.0 nm, and loss parameter $\gamma$ is 0.17 eV. Both magenta and blue lines are calculated from self-consistent hydrodynamics model, but the setting of loss parameter is different. For magenta lines, the loss parameter $\gamma$ is a constant throughout the domain $\Omega_D$, while for the blue lines, the loss parameter $\gamma$ is nonzero only within the jellium domain, i.e. setting to a nonzero constant in the jellium domain and zero outside the jellium.

Another important issue is about the nonlocal effect. Usually the HW-HDM is used to include the nonlocal response of metallic particles. To demonstrate this, we plot imaginary parts of $\mathbf{J}_1 / \mathbf{E}_1$ as a function of position for the dipole mode in Fig. S6(a). The gray, blue, and red lines are calculated by LRA, HW-HDM, and SC-HDM, respectively. The deviations of blue and red lines from gray line (which is by definition local) indicate that the nonlocal effect exists at that position, and both HW-HDM and SC-HDM indicate that nonlocal response is important near the metal surface. However, nonlocal effect only exist within the metal region for HW-HDM, while for SC-HDM, the nonlocal effect also exist in the spill out region. SC-HDM is of course closer to reality. For comparison, we also plot the imaginary parts of $\mathbf{J}_1 / \mathbf{E}_1$ for the Bennett mode by SC-HDM, as shown by red line in Fig. S6(b). We see that nonlocal effect is prominent near the metal surface.

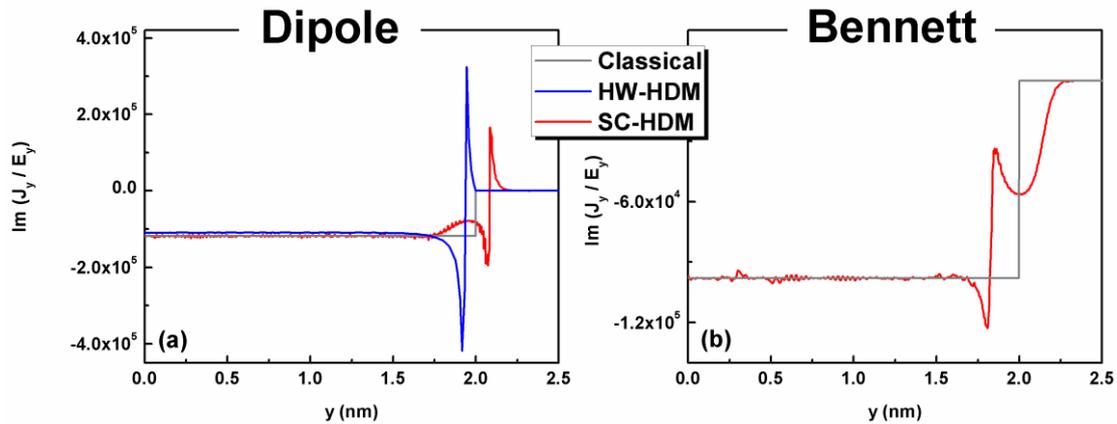

**Figure S6.** The imaginary parts of $\mathbf{J}_1 / \mathbf{E}_1$ as a function of radial coordinate for the dipole mode and Bennett mode claimed in Fig. S5 are plotted in (a) and (b), respectively. The solid gray lines, blue lines and red lines are obtained by classical model (LRA), HW-HDM, and SC-HDM, respectively.

## Section VII. Analytical derivations of Feibelman parameter for cylindrical particles

The full expressions of Feibelman d-parameters for cylindrical particles are [19]

$$\left(\frac{\omega}{\omega_p}\right)^2 = \frac{1}{2}\left(1 + \Lambda^{(0)} + d_\perp \Lambda_\perp + d_{//} \Lambda_{//}\right), \tag{S7.1}$$

where

$$\Lambda^{(0)} = \tilde{k}\left[K_m(\tilde{k}) I_m(\tilde{k})\right]' \qquad \Lambda_\perp = \frac{(\Lambda^{(0)})^2 - 1}{2 K_m(\tilde{k}) I_m(\tilde{k}) r_a}, \tag{S7.2}$$

$$\Lambda_{//} = 2 K_m(\tilde{k}) I_m(\tilde{k}) \frac{m^2 + \tilde{k}^2}{r_a} \qquad \tilde{k} = k r_a, \tag{S7.3}$$

$k$ is the wavenumber in radial direction. $K_m$ and $I_m$ are modified Bessel functions. Since we are concerned with the dipole resonance for subwavelength particles, so we take the limit $m > 0$ and $k r_a \to 0$. At this limit, we could evaluate Eqs. (S7.2) and (S7.3) as

$$\Lambda^{(0)} = 0 \qquad \Lambda_\perp = -\frac{m}{r_a} \qquad \Lambda_{//} = \frac{m}{r_a}. \tag{S7.4}$$

Substituting Eq. (S7.4) into Eq. (S7.1) gets the equation (4.1) used in the main text.

## Section VIII. Analytical derivations of the edge mode dispersions

Since the particles considered in this work are deep subwavelength, so we aim to analytically derive the dispersion relations of the edge modes (corner mode in 2D) within the quasi-static limit in this section. The method here follows the derivations in Ref. [20], with some extensions. We consider the configurations of a single metallic corner as shown in Fig. S7, and there are two media in the system. One is within the angle $0 < \theta < 2\alpha$ ($\alpha < \pi$), and the other exists within $2\alpha < \theta < 2\pi$. For simplicity, we assume the medium within $(0, 2\alpha)$ to be metallic with the permittivity $\varepsilon(\omega)$, and the medium outside this wedge is vacuum.

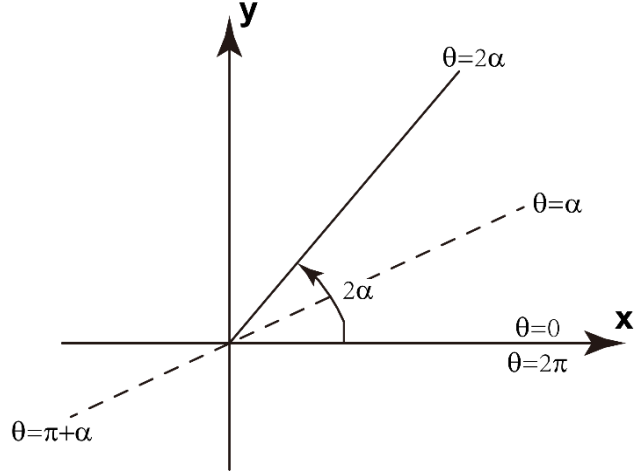

**Figure S7.** Schematic picture of the wedge configurations.

Since we only consider quasi-statics here, so we solve the electrostatic potential $\varphi$, satisfying the following Laplace's equation

$$\nabla^2 \varphi = 0, \tag{S8.1}$$

within each medium domain. Next, we separate variables by setting

$$\varphi(r,\theta) = R(r)\Theta(\theta), \tag{S8.2}$$

and then obtain the following pair of equations in the polar coordinates

$$r\frac{d}{dr}\left(r\frac{dR}{dr}\right) - m^2 R = 0, \tag{S8.3}$$

$$\frac{d^2}{d\theta^2}\Theta = -m^2\Theta, \tag{S8.4}$$

where $m$ is the separation constant. Two independent solutions of Eq. (S8.3) are $\{r^m, r^{-m}\}$ for $m \neq 0$ and $\{\ln r, \text{const.}\}$ for $m = 0$. Since we focus on the localized modes, we only keep the decaying solutions in the radial directions, denoted by $\tilde{R}_m(r)$. Two independent solutions of Eq. (S8.4) are $e^{\pm im\theta}$. Since we consider the mode localized near the wedges, so $m$ should be a purely imaginary number as $i\nu$. Furthermore, we could choose the two independent solutions in the angle direction as $\{\cosh(\nu\theta), \sinh(\nu\theta)\}$. Another point is that the system being studied has the reflection symmetry about the planes $\theta = \alpha$ and $\theta = \pi + \alpha$, we choose our solutions of Eqs. (S8.3) and (S8.4) to be even or odd about these planes, namely

$$\varphi^{(e)}(r,\theta) = \begin{cases} A\, \tilde{R}_{i\nu}(r)\cosh[\nu(\theta-\alpha)], & 0<\theta<2\alpha \\ B\, \tilde{R}_{i\nu}(r)\cosh[\nu(\theta-\alpha-\pi)], & 2\alpha<\theta<2\pi \end{cases} \quad (S8.5)$$

$$\varphi^{(o)}(r,\theta) = \begin{cases} C\, \tilde{R}_{i\nu}(r)\sinh[\nu(\theta-\alpha)], & 0<\theta<2\alpha \\ D\, \tilde{R}_{i\nu}(r)\sinh[\nu(\theta-\alpha-\pi)], & 2\alpha<\theta<2\pi \end{cases} \quad (S8.6)$$

where the superscript "e" (o) denotes the even (odd) mode. The boundary conditions of Eq. (S8.1) are the potential $\varphi$ and the normal component of the displacement field, $\mathbf{D}_\theta = -\varepsilon_0 \varepsilon_r r^{-1}(\partial\varphi/\partial\theta)$, should be continuous at the boundaries $\theta = 2\alpha$ and $\theta = 0$. For the even mode, the continuity of $\varphi$ at $\theta = 2\alpha$ gives out

$$\frac{A}{B} = \frac{\cosh[\nu(\pi-\alpha)]}{\cosh(\nu\alpha)}, \quad (S8.7)$$

and the continuity of $\mathbf{D}_\theta$ at $\theta = 2\alpha$ gives out

$$\frac{A}{B} = -\frac{\sinh[\nu(\pi-\alpha)]}{\varepsilon(\omega)\sinh(\nu\alpha)}. \quad (S8.8)$$

Combine Eqs. (S8.7) and (S8.8) will get the following dispersion relation for the even mode

$$\varepsilon(\omega) = -\frac{\tanh[\nu(\pi-\alpha)]}{\tanh(\nu\alpha)}. \quad (S8.9)$$

It could be shown that the boundary conditions at $\theta = 0$ are satisfied automatically after matching the boundary conditions at $\theta = 2\alpha$. Similarly, the dispersion relations of the odd mode could also be obtained as

$$\varepsilon(\omega) = -\frac{\tanh(\nu\alpha)}{\tanh[\nu(\pi-\alpha)]}. \quad (S8.10)$$

It could be seen that the even mode is at the lower frequencies, while the odd mode is at the higher frequencies. So the fundamental edge mode in the main text should be the even modes. To see the dispersions of this mode, we take $\nu \to 0$, and then Eq. (S8.9) gives out the angle dependence of the edge mode as

$$\left(\frac{\omega}{\omega_p}\right)^2 = \frac{\alpha}{\pi}. \tag{S8.11}$$

This is the equation (4.3) used in the main text. Furthermore, the dispersions (S8.11) are also valid if we smooth the corner singularities [21].

**Section IX. Discussion of the boundary conditions for charged particles**

In this section, we will show the choice of the ground state calculation domain $\Omega_G$ for charged particles. As mentioned in the Sec. II, the values of $\nabla \phi_0' / k_{\mathrm{TF}}^2$ on the boundary depend on the size of $\Omega_G$ for the same amount of additional charges. So for the same physical system, changing the calculation domain size of $\Omega_G$ also modifies the boundary conditions applied. To verify that this artificial setting does NOT affect the results, we first calculate the ground state results for a charged triangle, as shown in Figs. S8(a) and S8(b). The red dots, green line, and blue line are obtained by setting the radius of $\Omega_G$ to 5nm, 8nm, and 10nm, respectively. It can be seen that they have the same results in their common regions. We also calculate the absorption spectra and optical forces of this charged triangle for these different choice of $\Omega_G$, as shown in Figs. S8(c) and S8(d). The results also confirm that they give out the same result. By recognizing this, we set $r_g$ to 8.0 nm throughout this work for charged particle calculations.

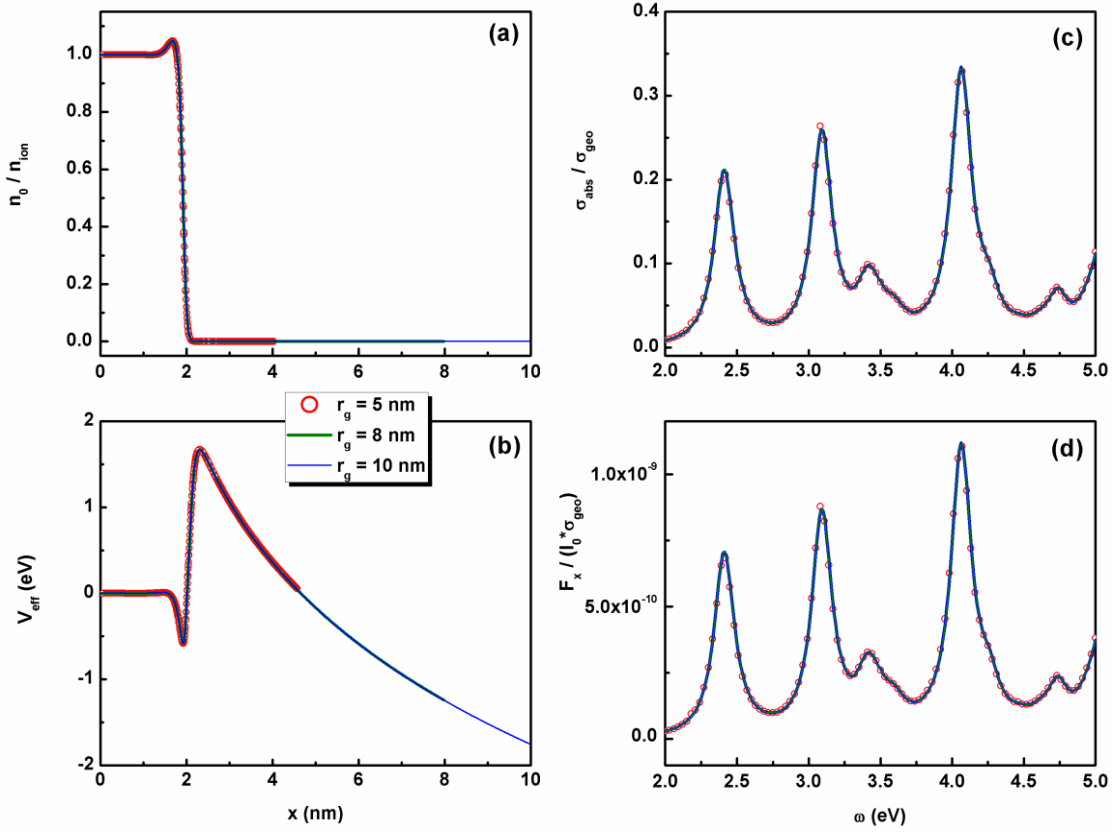

**Figure S8.** Electron distributions $n_0/n_{ion}$ and effective potential $V_{eff}$ for a charged sodium triangle ($\Delta S = -0.01\pi$ nm$^2$) are plotted in (a) and (b), respectively, for different values of ground state calculation domain radius $r_g$. Absorption spectra and total optical force of this charged sodium triangle are shown in (c) and (d), respectively, for different values of $r_g$.

We can then calculate the ground state results for different values of $\sigma$. In order to make the doping charges $\sigma$ easier to interpret, we introduce the quantity $\Delta S$ defined as $\sigma/en_{ion}$ with the dimension of area in 2D cases here. $\Delta S > 0$ ($<0$) stands for p-charging (n-charging) particles. The ratio of $\Delta S$ to the total jellium area means the percentage of number of charge carriers (dopant) to the total number of charge carriers. The calculated electron density $n_0/n_{ion}$ and effective potential $V_{eff}$ as a function of position in the E-direction of triangles for different charging cases are plotted in Figs. S9(a) and S9(b), respectively. We see that the electrons move toward the interior of jellium for p-charging cases, while the electrons shift outwards for n-charging cases. Accordingly, the effective potentials are also shifted inwards or outwards for p-charging and n-charging cases. Furthermore, the $n_0/n_{ion}$ and $V_{eff}$ in the F-direction for triangles for different charging cases are also plotted in Figs. S9(c) and S9(d), respectively. We see that the shifts caused by the additional charges are smaller in the F-direction than that in the E-direction. This means that the additional charge carriers tend to accumulate in the sharper edges than in the smoother faces. This trend also leads to the change of the surface static dipoles which will provide additional screening in linear responses.

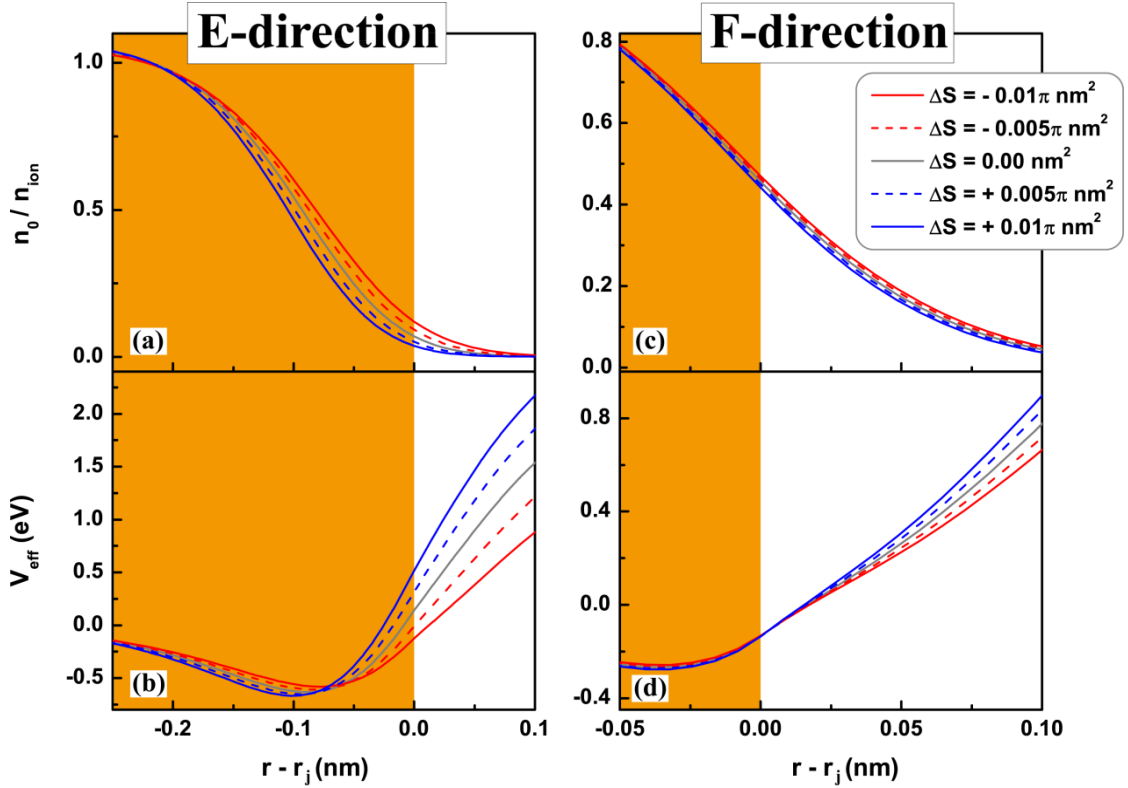

**Figure S9.** Electron distributions $n_0/n_{ion}$ and effective potential $V_{eff}$ in the E(edge)-direction for a charged sodium triangle are shown in (a) and (b), respectively, for different additional amounts of $\Delta S$. The $n_0/n_{ion}$ and $V_{eff}$ in the F(face)-direction for the charged sodium triangle are plotted in (c) and (d), respectively. The radii of these triangles are all 2.0 nm. The zero of horizontal axis is set to the jellium edge, and the yellow region stands for the jellium domain.

**Section X. Energy dissipation distributions**

We plotted the energy dissipation distributions of different plasmonic modes in the main text, so one benchmark calculation worthy to carry out is to check the following equality

$$\int_{\Omega_D} d\mathbf{r} \langle \mathbf{J}_1 \cdot \mathbf{E}_1 \rangle = -\int_{\partial \Omega_G} d\mathbf{A} \cdot \langle \mathbf{S} \rangle, \quad (S10.1)$$

where $\langle \rangle$ denotes time average, and $\mathbf{S}$ is the Poynting vector. We plot the left side and right side results of Eq. (S10.1) for a charge neutral triangle in Fig. S10 by blue dots and red lines, respectively. The agreement between them confirms Eq. (S10.1) is valid in our model.

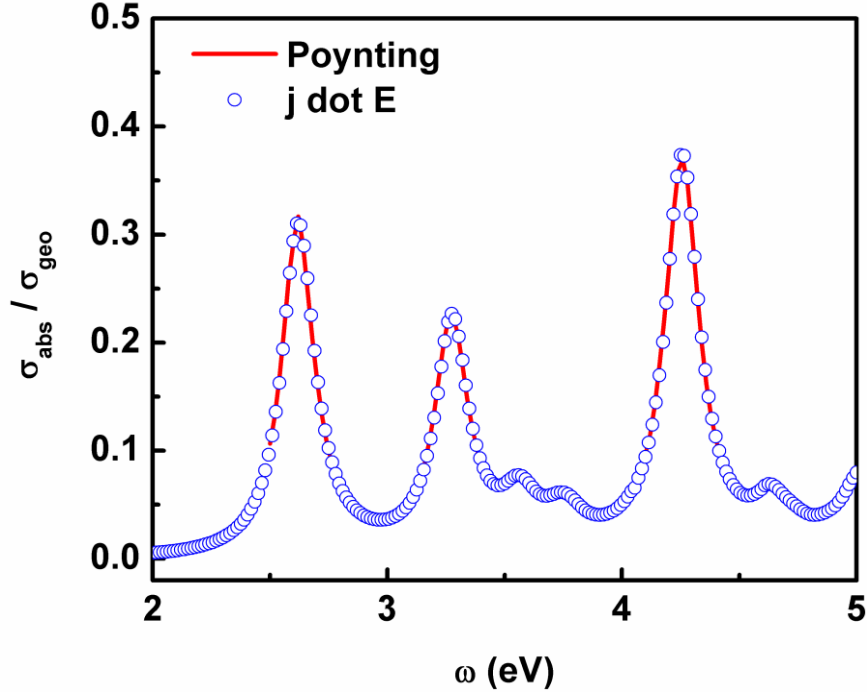

**Figure S10.** Absorption spectra of a charge neutral triangle with circumradius 2.0 nm are plotted, in which solid red lines are calculated by surface integral of Poynting vector in the far field, and open blue circles are obtained by volume integral of $\langle \mathbf{J}_1 \cdot \mathbf{E}_1 \rangle / I_0$ in the domain $\Omega_D$.